\documentclass[sort&compress,AMA,STIX1COL]{WileyNJD-v2}

\articletype{Research Article}%

\received{26 April 2016}
\revised{6 June 2016}
\accepted{6 June 2016}

\usepackage{tikz}
\usetikzlibrary{arrows}
\usetikzlibrary{shapes}

\newcolumntype{P}[1]{>{\centering\arraybackslash}p{#1}}
\newcolumntype{M}[1]{>{\centering\arraybackslash}m{#1}}

\makeatother

\usepackage[capitalise]{cleveref}	
	\crefformat{equation}{(#2#1#3)}	
	
\usepackage{amsmath}
\usepackage{pgfplots}
\usepackage{float}
\usepackage{graphicx}
\usepackage{siunitx}
\usepackage{caption}
\usepackage{subcaption}
\usepackage{comment}
\usepackage{pbox}
\definecolor{myGreen}{RGB}{0, 176, 80}
\definecolor{myRed}{RGB}{254, 0, 0}

\usepackage{soul}
\usepackage{hyperref}
\usepackage{bm}

\definecolor{blue}{RGB}{0, 0, 0}\definecolor{myGreen}{RGB}{0, 0, 0}\definecolor{myRed}{RGB}{0, 0, 0}

\pgfplotsset{width=7cm,compat=1.8}
\begin{document}

\title{Energy-aware Relay Positioning in Flying Networks}

\author{Hugo Rodrigues}

\author{André Coelho}

\author{Manuel Ricardo}

\author{Rui Campos}

\authormark{Hugo Rodrigues \textsc{et al.}}

\address{INESC TEC and Faculdade de Engenharia, Universidade do Porto, Portugal.}

\corres{INESC TEC and Faculdade de Engenharia, Universidade do Porto, 4200-465 Porto, Portugal.
E-mail addresses: rodrigues97hugo@gmail.com (Hugo Rodrigues); andre.f.coelho@inesctec.pt (André Coelho); manuel.ricardo@inesctec.pt (Manuel Ricardo); rui.l.campos@inesctec.pt (Rui Campos).}

\abstract[Summary]{The ability to move and hover has made rotary-wing Unmanned Aerial Vehicles (UAVs) suitable platforms to act as Flying Communications Relays (FCR), aiming at providing on-demand, temporary wireless connectivity when there is no network infrastructure available or a need to reinforce the capacity of existing networks. However, since UAVs rely on their on-board batteries, which can be drained quickly, they typically need to land frequently for recharging or replacing them, limiting their endurance and the flying network availability. The problem is exacerbated when a single FCR UAV is used. The FCR UAV energy is used for two main tasks: communications and propulsion. The literature has been focused on optimizing both the flying network performance and energy-efficiency from the communications point of view, overlooking the energy spent for the UAV propulsion. Yet, the energy spent for communications is typically negligible when compared with the energy spent for the UAV propulsion.

In this article we propose Energy-aware RElay Positioning (EREP), an algorithm for positioning the FCR taking into account the energy spent for the UAV propulsion. Building upon the conclusion that hovering is not the most energy-efficient state, EREP defines the trajectory and speed that minimize the energy spent by the FCR UAV on propulsion, without compromising \textcolor{blue}{in practice} the Quality of Service offered by the flying network. The EREP algorithm is evaluated using simulations. The obtained results \textcolor{blue}{show gains up to 26\% in the FCR UAV endurance for negligible throughput and delay degradation.}}

\keywords{Aerial Networks, Energy-aware, Flying Networks, Relay Positioning, Unmanned Aerial Vehicles, UAV trajectory.}

\maketitle

\section{Introduction}\label{sec:Introduction}
\begin{figure}
	\centering
	\includegraphics[scale=0.20]{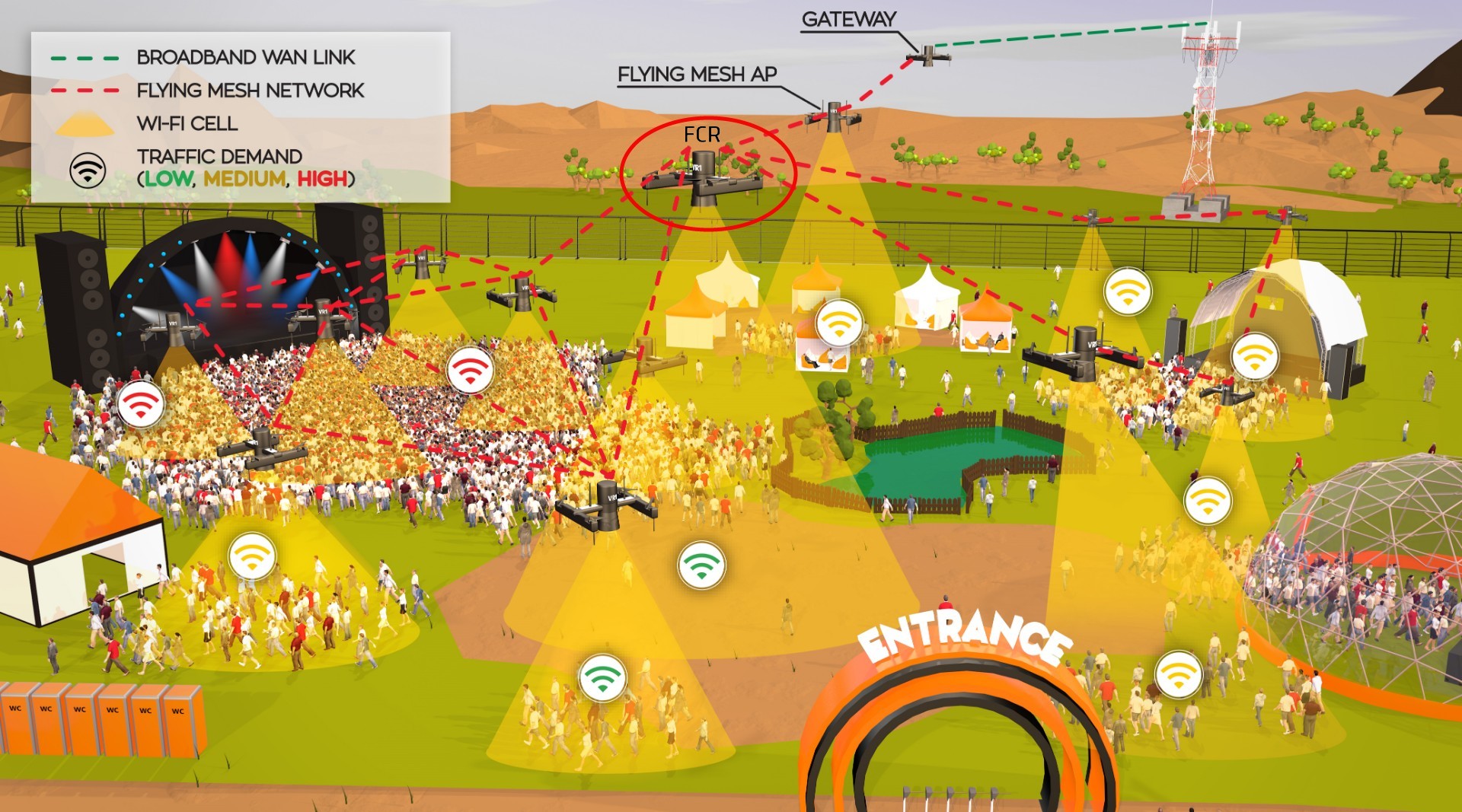}
	\caption{Flying Network providing Internet connectivity to the users in a music festival. The FCR UAV is circled in red\textcolor{blue}{\cite{coelho2018redefine}}.}
	\label{fig:Music Festival Scenario}
\end{figure}

In recent years there has been an increase in the usage of Unmanned Aerial Vehicles (UAVs) for a myriad of applications \cite{motlagh2017uav}. Their capability to operate almost everywhere, their ability to hover, and their increasing capacity to carry cargo on board make UAVs suitable platforms to act as Flying Communications Relays (FCRs). This reality has prompted the interest in using flying networks to establish and reinforce communications and enable broadband Internet access in temporary events \cite{coelho2018redefine, almeida2018traffic}, as depicted in \cref{fig:Music Festival Scenario}. 
A major challenge in flying networks when compared with ground networks is the UAV endurance. As the UAVs are not connected to the electrical grid, they rely on their on-board batteries, which can be drained quickly. For this reason, the UAVs typically need to land frequently for recharging or replacing their batteries, which limits the UAV endurance and the flying network availability \cite{zeng2016wireless}. The problem is exacerbated if the UAV plays the role of FCR UAV, especially in a flying network composed of a single FCR. Since the FCR UAV is responsible for forwarding the traffic to/from the Internet, its available energy directly affects the Quality of Service (QoS) offered by the flying network. If the FCR UAV becomes unavailable due to energy shortage, the rest of the network will be unable to connect to the Internet. 

Energy in flying networks composed of UAVs is used for two main tasks: communications and UAV propulsion \cite{Alzahrani2020}. While the energy spent for communications is typically negligible, the energy spent for propulsion should be taken into account when defining the positioning of the UAVs. In this article, we assume the flying network is composed of rotary-wing UAVs, which are divided into two types: Flying Access Points (FAPs) and FCR UAVs. Solutions have been proposed in the literature to define the positioning and trajectories of 1) FAPs, considering  the  radio coverage, the  number of ground users served \cite{Mozaffari2015, Kalantari2017, Wu2018, Arribas2019} or the QoS offered to the ground users \cite{Zeng2016, Alzenad2017, almeida2018traffic, Almeida2019, Liu2019, Almeida2020}, and 2) UAVs acting as relays between ground nodes \cite{larsen2017optimal, Zhong2019, Zhong2020}. However, most of the works have been focused on optimizing both the flying network performance and energy-efficiency from the communications point of view \cite{Li2015, Li2016, Li2019, Qin2019trajectory}, overlooking the energy spent for the UAV propulsion.

The main contribution of this article is the Energy-aware RElay Positioning algorithm (EREP) for flying networks with controlled topology. Building upon the conclusion that hovering is not the most power-efficient state \cite{zeng2019energy}, EREP defines the trajectory and speed of a rotary-wing FCR UAV that minimizes its energy consumption for propulsion, without compromising the QoS offered by the flying network. EREP represents a step forward with respect to the algorithm proposed by Coelho et al. \cite{coelho2019traffic}, which defines the positioning of an FCR UAV from the network performance point of view, without considering the energy consumption of the UAV.

The rest of this article is organized as follows.
Section \ref{sec:RelatedWork} presents the related work.
Section \ref{sec:ProblemFomulation} presents the system model and formulates the problem.
Section \ref{sec:ProposedAlgorithm} presents the EREP algorithm.
Section \ref{sec:PerformanceEvaluation} presents the evaluation of the EREP algorithm using simulations.
Finally, Section \ref{sec:Conclusions} points out the main conclusions and future work.

\section{Related Work}\label{sec:RelatedWork}
In order to ensure energy-efficiency in communications, the literature has been focused on optimizing the performance of different communications layers: 1) network layer, by designing energy-aware routing protocols that define the forwarding tables based on the energy available in the UAVs \cite{Oubbati2019,Aadil2018,Alshabtat2010}; 2) data link layer, by enhancing the MAC scheme, including the usage of sleep modes on idle communications nodes \cite{Fan2012, Yang2004}; and 3) physical layer, by optimizing the hardware and the communications tasks related to the signal transmission and processing\cite{Mozaffari2017}. Nevertheless, the energy used in communications is typically negligible when compared with the energy spent for the UAV propulsion. For instance, a typical IEEE 802.11n Network Interface Card has a maximum power consumption of about \SI{2}{\watt} \cite{Halperin2010, Sousa2018}, while the propulsion power consumption of a UAV in hovering state can be higher than \SI{150}{\watt}, as depicted in \cref{fig:power}. A reference work on joint energy-efficiency and trajectory optimization for an FCR UAV was done by Zeng et al.\cite{Zeng2017}. The authors derived a theoretical model for estimating the energy spent for the UAV propulsion, based on the speed, direction, and acceleration of the UAV, and proposed a circular trajectory in which the radius and the flight speed are optimized to maximize the energy efficiency -- defined as \SI{}{bit/\joule}. However, the proposed model is tailored for fixed-wing UAVs and considers a single ground user only. Li et al. \cite{Xiaowei2019, Xiaowei2019IoT} proposed an energy-efficiency model for a multi-UAV flying network, including components regarding the UAVs' energy consumption and the deployment strategy for recharging the UAVs, in order to provide seamless long-term coverage. However, the UAVs are positioned in hovering state to provide communications cells and their movement to/from the charging station is not performed at the speed that minimizes their energy consumption.

When it comes to the FCR UAV positioning challenge, Coelho et al. \cite{coelho2019traffic} proposed a traffic-aware positioning algorithm for flying networks with controlled topology. The proposed algorithm considers the traffic generated by each of the FAPs to define the position of the FCR, which acts as a flying gateway between the FAPs and the Internet. However, the proposed algorithm neglects the energy consumption for the UAV propulsion. A reference work on energy-efficiency for rotary-wing UAVs was done by Zeng et al. \cite{zeng2019energy}, who have concluded that the power consumption does not have a uniform behaviour: the power consumption for low speed values is lower than the power consumption for hovering and it becomes higher as the speed increases. Therefore, hovering is not the most power-efficient UAV state. 
In order to improve energy-efficiency, the authors propose a fly-hover approach, in which the FCR UAV moves between different waypoints, where it hovers for a period of time for providing communications for each ground user, in a Time-Division Multiple Access scheme.
Demir et al. \cite{demir2019energy} studied the power-aware deployment of a UAV, considering flight dynamics and QoS guarantees for the users being served. Still, their work is focused on the influence of the altitude, hardware components, and payload weight in the UAV power consumption, considering the UAV is hovering. Wang et al. \cite{Jingjing2019} proposed a two-stage joint hovering altitude and power control model for space-air-ground communications using UAVs. A related research work was carried out by Babu et al. \cite{Babu2020_1, Babu2020_2}, who derived the optimal altitude for rotary-wing UAVs acting as FAPs, taking into account both the UAVs energy consumption and the communications performance. Nevertheless, these works consider the UAVs in hovering state only.

\section{System Model and Problem Formulation}\label{sec:ProblemFomulation}

\begin{figure}
	\centering
	\includegraphics[scale=0.5]{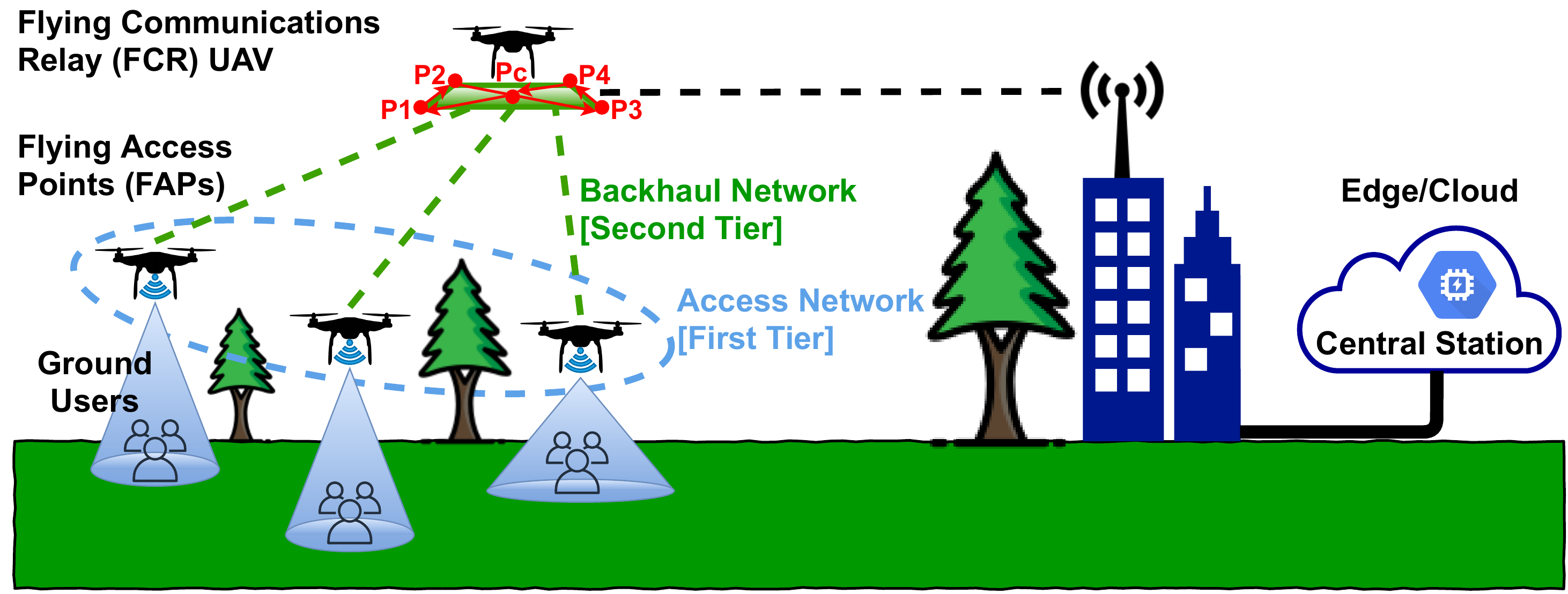}
	\caption{Flying network organized into a two-tier architecture. The FAPs provide Internet access to the ground users and the FCR UAV forwards the traffic to/from the Internet. The flying network is controlled by a Central Station (CS) deployed at the Edge or in the Cloud. An example trajectory defined by the EREP algorithm is depicted by the red arrows\textcolor{blue}{\cite{coelho2019traffic}}.}
	\label{fig:Flying network architecture}
\end{figure}

The flying network, which is assumed to be organized into a two-tier architecture, is composed of two types of UAVs: 1) FAPs, which provide Internet access to ground nodes; 2) a single FCR UAV, which forwards the traffic to/from the Internet, as depicted in \cref{fig:Flying network architecture}. The positions of the FAPs are assumed to be defined by a state of the art FAP positioning algorithm, such as the centralized algorithm proposed by Almeida et al.\cite{almeida2018traffic}. The FAP positioning algorithm is assumed to run on a Central Station (CS). The CS periodically determines the coordinates of the FAPs based on the traffic demand of the ground nodes, which is collected by the FAPs themselves and transmitted to the CS. The FCR UAV positioning is defined by the CS considering the current FAPs’ coordinates and traffic demand. Finally, the CS is in charge of sending the updated positions to both the FAPs and the FCR UAV, which position themselves accordingly.

In the following, we formulate the problem addressed in this article. To calculate the received power, as there is Line-of-Sight (LoS) between the FCR UAV and the FAPs, the Free-Space Path Loss model defined in \cref{eq:free_space_path_loss_model}\textcolor{blue}{\cite{friis1946note}} is considered. This assumption is supported by Khuwaja et al. \cite{Khuwaja2018} and Almeida et al. \cite{Almeida2020}, who have concluded, based on experimental results, that Free-Space Path Loss model is the most adequate to characterize the wireless link between a UAV and a communications node close to the ground in terms of the path loss component.

\begin{equation}
    \label{eq:free_space_path_loss_model}
    \frac{P_r}{P_t} = \left[\frac{\lambda}{4\pi d}\right] ^2 
\end{equation}

In \cref{eq:free_space_path_loss_model}, ${P_r}$ stands for the power received at the FCR UAV, ${P_t}$ is the transmission power of each FAP, the wavelength $\lambda$ is equal to $c/f$, where ${c}$ is the speed of light in vacuum and ${f}$ is the carrier frequency, \textcolor{blue}{$\pi$ is the mathematical constant defined in Euclidean geometry as the ratio of a circle's circumference to its diameter,} and $d$ represents the distance between the transmitter and receiver UAVs. We assume that the maximum channel capacity is equal to the data rate associated to the Modulation and Coding Scheme (MCS) index selected by the nodes. Each MCS index requires a minimum value of Signal-to-Noise Ratio, $SNR=P_r/N_0$, which is derived from $P_r$ considering a constant noise power $ N_0$. The wireless medium is shared, so we assume that every UAV can listen to the other UAVs. For Medium Access Control (MAC) the Carrier Sense Multiple Access with Collision Avoidance (CSMA/CA) is employed. 

The power consumed by the UAV for propulsion was defined by Zeng et al. \cite{zeng2019energy} as having three components: 1) blade profile, which is the power required to overcome the profile drag of the blades; 2) induced, which is required to overcome the induced drag of the blades; and 3) parasite, required to overcome the fuselage drag. As such, while the blade profile power increases quadratically with $V$, the parasite power increases cubically with $V$. On the other hand, the induced power decreases with $V$. The equation for calculating the power $P$ consumed by the UAV while moving at speed $V$ is given in \cref{eq:power-consumption}\textcolor{blue}{\cite{zeng2019energy}}.

\begin{equation}
    \begin{split}
        P(V)= & P_b\left(1+\frac{3V^2}{U^2_{tip}}\right)+P_{ind}\left(\sqrt{1+\frac{V^4}{4v^4_0}}-\frac{V^2}{2v^2_0}\right)^{3/2} + \frac{1}{2}d_0 \rho s AV^3
    \end{split}
    \label{eq:power-consumption}
\end{equation}

\begin{figure}
	\centering
	\includegraphics[width=0.4\linewidth]{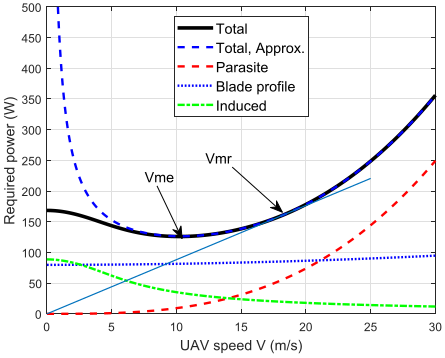}
	\caption{Propulsion power consumption versus UAV speed $V$ \cite{zeng2019energy}.}
	\label{fig:power}
\end{figure}

In \cref{eq:power-consumption}, the first addend represents the blade profile component, where ${P_b}$ is a constant representing the blade profile power in hovering state, and ${U_{tip}}$ denotes the tip speed of the rotor blade. The second addend represents the induced power component, where ${P_{ind}}$ is a constant representing the induced power in hovering state, and ${v_0}$ is the mean rotor induced velocity in hovering state. The third addend represents the parasite component, where ${d_0}$ is the fuselage drag ratio, \textcolor{myGreen}{${\rho}$ denotes the air density,} ${s}$ represents the rotor solidity, and ${A}$ represents the rotor disc area. These parameters can be obtained from the UAV specifications, with the exception of ${\rho}$, whose value depends on the environment. The analysis of \cref{eq:power-consumption} made by Zeng et al. \cite{zeng2019energy} shows that there is a range of UAV speeds $V$ for which the power consumed by the UAV is lower than the power consumed for hovering. By defining $V=0$ in~\cref{eq:power-consumption}, we conclude that the power consumption in hovering state is equal to $P_b + P_{ind}$, which represents a finite value that changes according to factors such as the UAV weight, air density, and rotor disc area. When $V$ increases, $P(V)$ firstly decreases and then increases with $V$. This allows to conclude that hovering is not the most efficient state when it comes to the UAV power consumption. This behavior is depicted in \cref{fig:power}.

In our problem, the flying network is modeled as a directed graph $G=(U, L)$ where $U=\{UAV_0, ..., UAV_{N-1} \}$ is the set of UAVs $i$ positioned at $Q_i=(x_i, y_i, z_i)$ and $L \subseteq  U \times U $ is the set of directional links between UAVs $i$ and $j$, with $(i,j)\in L$ and $i, j \in U$. Let us assume that each $UAV_i$, $i\in \{1, ..., N-1\}$, performs the role of FAP and transmits a traffic flow $F_{0,i}$ towards $UAV_0$, which performs the role of FCR UAV. In this sense, we have a tree $T(U, L_T)$ that is a subgraph of $G$, where $L_T \subset L$ is the set of direct links between each $UAV_i$ and $UAV_0$. This tree defines the flying network active topology, which is composed of single-hop paths.

Considering $T$ the $UAV_0$ endurance and $0\leq t \leq T$, we aim at determining the trajectory $Q_0(t) = \left (x_0(t),y_0(t),z_0(t) \right )$ of $UAV_0$, such that the power $P_0\left (\left \| \dot{Q_0}(t)  \right \|\right )$ consumed by $UAV_0$ for propulsion is minimal, where $\left \| \dot{Q_0}(t)  \right \|$ represents the $UAV_0$ speed at time instant $t$, and the transfer of all traffic flows $F_{0,i}$ with bitrate $T_i(t)$, in bit/s, is guaranteed, taking into account the maximum power $P_0^{MAX}$ and the maximum speed $V_0^{MAX}$ allowed by $UAV_0$. For that purpose, a wireless link with high enough capacity $C_{0,i}(t)$ to accommodate $T_i(t)$ must be ensured between each $UAV_i$, $i \in \{1, ..., N-1\}$, and $UAV_0$, while the aggregate capacity of the wireless links must be lower than or equal to the maximum capacity $C^{MAX}$ of the wireless channel. This is accomplished by ensuring the minimum $SNR_i$ at $UAV_i$ that enables the selection of an MCS index characterized by a data rate higher than or equal to $T_i(t)$ bit/s, which depends on the Euclidean distance between each $UAV_i$ and $UAV_0$, and the transmission power $P_{t}$ of the UAVs, whose maximum value $P_{t}^{MAX}$ is defined based on the wireless technology used. Our objective function is defined in \cref{eq:objective-function}.

\begin{subequations}
    \label{eq:optimization-problem}
	\begin{alignat}{4}
		  & \!\underset{T,\; Q_0(t)\,=\,\left (x_0(t),\: y_0(t),\: z_0(t)\right ), \;P_t}{\text{minimize}} & \; \; \; & \int_{0}^{T} P_0\left (\left \| \dot{Q_0}(t)  \right \|\right ) dt\label{eq:objective-function}\\ 
		  & \text{subject to:} & & 0 < P_0\left (\left \| \dot{Q_0}(t)  \right \|\right ) \leq P_0^{MAX}, \forall t \in \left [ 0, T \right] \label{eq:constraint1}\\ %
		  & & & 0 \leq P_t \leq P_t^{MAX} \label{eq:constraint2}\\
		  & & & \left \| \dot{Q_0}(t)  \right \| \leq V_0^{MAX}, \forall t \in \left [ 0, T \right]\label{eq:constraint3}\\
		  & & & Q_0(t) \neq (x_i, y_i, z_i), \forall t \in \left [ 0, T \right], i \in \{1,...,N-1\} \wedge  z_i \geq 0, i \in \{0,...,N-1\}\label{eq:constraint4}\\
		  & & & (0, i), (i, 0) \in L_T, i \in \{1, ..., N-1\} \label{eq:constraint6}\\
          & & & 0 < T_i(t) \leq C_{0,i}(t), i \in \{1,...,N-1\}, \forall t \in \left [ 0, T \right]\label{eq:constraint7}\\
          & & & \sum_{i=1}^{N-1}C_{0,i}(t) \leq C^{MAX}, i \in\{1,...,N-1\}, \forall t \in \left [ 0, T \right]\label{eq:constraint8} \\
		  & & & \left(x_0(t) \smash{-} x_i\right)^2 + \left(y_0(t) \smash{-} y_i\right)^2 + \left(z_0(t) \smash{-}  z_i\right)^2 \leq \left(10^{\frac{P_t \smash{-} 20\log_{10}\left(\frac{4\pi}{\lambda}\right) \smash{-} N_0 \smash{-}  SNR_i}{20}}\right)^2, \forall t \in \left [ 0, T \right], i \in \{1, ..., N-1\}\label{eq:constraint9}
	\end{alignat}
\end{subequations}

In the defined optimization problem, the following constraints are considered for any time instant $t$: 
\begin{itemize}
    \item \cref{eq:constraint1} ensures that the propulsion power consumption is higher than 0, in order to enable the operation of $UAV_0$, and lower than or equal to the maximum power $P_0^{MAX}$ supported by $UAV_0$. \item \cref{eq:constraint2} guarantees that the transmission power $P_{t}$ is between 0 and the maximum value $P_{t}^{MAX}$ defined based on the wireless cards used. 
    \item \cref{eq:constraint3} assures that the speed of $UAV_0$ at any time instant $t$ is lower than or equal to the maximum speed $V_0^{MAX}$ allowed by $UAV_0$. 
    \item \cref{eq:constraint4} ensures that the position of $UAV_0$, which we aim at determining, is different from any $UAV_i$ composing the flying network, in order to avoid collisions. We assume the position of any $UAV_i, i \in \{1,...,N-1\}$, is defined by a state of the art FAP positioning algorithm, such as the one proposed by Almeida et al.~\cite{almeida2018traffic}, which avoids collisions between them.
    \item \cref{eq:constraint6} ensures that a wireless link between each $UAV_i$ and $UAV_0$ is always available. 
    \item \cref{eq:constraint7} guarantees that the capacity of the wireless link established between each $UAV_i$ and $UAV_0$ is higher than or equal to the traffic demand of $UAV_i$. 
    \item \cref{eq:constraint8} assures that the aggregate capacity of the wireless links is lower than or equal to the maximum capacity $C^{MAX}$ of the wireless channel. 
    \item \cref{eq:constraint9} ensures the minimum $SNR_i$ that enables the selection of an MCS index characterized by a data rate higher than or equal to $T_i(t)$ bit/s, which depends on the Euclidean distance between each $UAV_i$ and $UAV_0$, and the transmission power $P_{t}$ of the UAVs, whose maximum value $P_{t}^{MAX}$ is defined based on the wireless technology used.
\end{itemize}

\cref{eq:objective-function} requires defining the trajectory of $UAV_0$ such that the propulsion energy consumption is minimal and the capacity of the wireless links established between the FAPs and $UAV_0$ is high enough to accommodate the traffic demand of each FAP. The UAV trajectory planning has been shown in the literature to be an NP-hard problem\cite{Zhan2018, Samir2019}. As such, we propose a heuristic algorithm to achieve a solution for this problem.

\section{Energy-aware Relay Positioning Algorithm}\label{sec:ProposedAlgorithm}
As previously mentioned, there is a range of UAV speeds wherein the UAV consumes less power than when it is hovering. This assumption is the basic principle for the Energy-aware RElay Positioning (EREP) algorithm proposed herein. The EREP algorithm is built upon the GWP algorithm proposed by Coelho et al.\cite{coelho2019traffic}, and takes into account the constraints included in the problem formulation presented in \cref{sec:ProblemFomulation}. GWP takes advantage of the knowledge of the FAPs positions and traffic demands, provided by a state of the art FAP positioning algorithm such as the one presented by Almeida et al.\cite{almeida2018traffic}. Then, it defines the position of the FCR UAV (named GW UAV) that maximizes the aggregated throughput between the FAPs and the FCR UAV. For that purpose, GWP computes a three-dimensional (3D) subspace that results from the intersection of the spheres centered at each FAP composing the flying network. Each sphere has a radius equal to the maximum distance that enables the minimum SNR value required for selecting a target MCS index. The target MCS index is characterized by a data rate higher than or equal to the traffic demand of the corresponding FAP.

\begin{table}
	\renewcommand{\arraystretch}{1.4}
	\def\arraystretch{1.8}
	\caption{Main differences between the EREP and GWP algorithms. While the GWP algorithm places the FCR UAV in hovering state, which is not the most energy-efficient UAV state according to the literature \cite{zeng2019energy}, the EREP algorithm defines a trajectory to be completed by the FCR UAV at the speed that minimizes the energy required for the UAV propulsion, without compromising the network performance.}
	\label{tab:differences-erep-gwp}
	\centering
	\begin{tabular}{|c|c|c|c|}
		\hline
		\pbox{20cm}{\textbf{Algorithm}} & \pbox{20cm}{\textbf{FCR UAV state during}\\\textbf{network operation}} & \pbox{20cm}{\textbf{Network performance}\\\textbf{awareness}} &  \pbox{20cm}{\textbf{Energy consumption}\\\textbf{for UAV propulsion}}\\ 
		\hline \hline
		EREP & Moving at constant speed & Yes & Minimum \\ \hline
		GWP & Hovering & Yes & High \\ \hline
	\end{tabular}
\end{table}

The EREP algorithm improves the GWP algorithm by considering the power consumption of the FCR UAV. Instead of hovering in the position defined by the GWP algorithm, with EREP the UAV is moved along a trajectory, within the 3D subspace, at the speed that minimizes the power consumption, given by \cref{eq:power-consumption}. Since any position within the 3D subspace allows to ensure the minimum SNR required for selecting the target MCS index, the trajectory defined by EREP for the FCR UAV does not compromise the QoS offered by the flying network when compared with the positioning defined by the GWP algorithm. The main differences between the EREP and GWP algorithms are presented in \cref{tab:differences-erep-gwp}. 

\begin{figure}[!ht]
	\centering
	\subfloat[Transmission range for each FAP.]{
		\includegraphics[height=0.2\linewidth,width=0.25\linewidth]{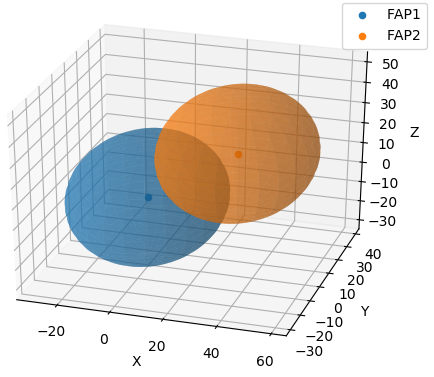}
		\label{fig:imageA}}
	\subfloat[Trajectory 1.]{
		\includegraphics[height=0.2\linewidth,width=0.25\linewidth]{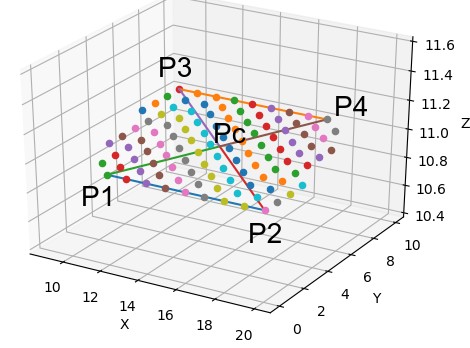}
		\label{fig:imageB}}
	\subfloat[Trajectory 2.]{
		\includegraphics[height=0.2\linewidth,width=0.25\linewidth]{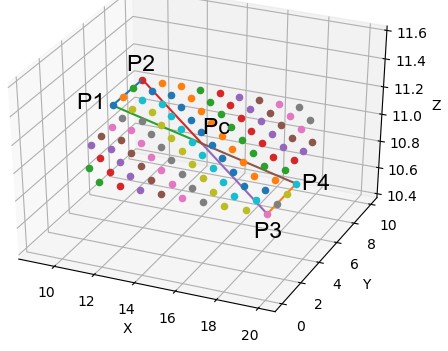}
		\label{fig:imageC}}
	\subfloat[Trajectory 3.]{
		\includegraphics[height=0.2\linewidth,width=0.25\linewidth]{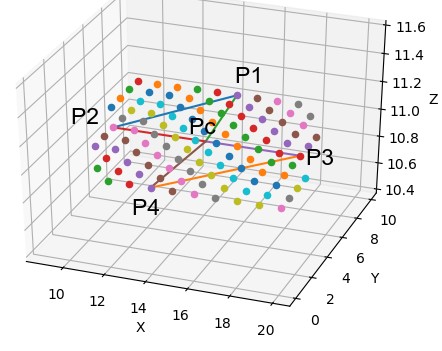}
		\label{fig:imageD}}
	\caption{Transmission range for each FAP and candidate FCR trajectories, considering a simplified networking scenario for illustrative purposes~\cite{rodrigues2020}.}
	\label{fig:set4images}
\end{figure}

The first step of the EREP algorithm consists in determining the minimum $SNR_i$ that enables the usage of an MCS index $MCS_i$ capable of accommodating the traffic demand $T_i(t)$ bit/s offered by $FAP_i$ (line 1 of Alg. 1). For that purpose, the relation between the SNR and the fair share of the wireless channel capacity is considered, following the rationale proposed by Coelho et al. \cite{coelho2019traffic}. For illustrative purposes, let us assume the use of the IEEE 802.11ac technology with one spatial stream, 800 ns Guard Interval (GI), and \SI{160}{\mega\hertz} channel bandwidth. Considering the minimum SNR required to use a target MCS index, which is characterized by a theoretical data rate, in \cref{tab:snr-fair-share} we calculate the fair share for 2 FAPs that use the same wireless channel, taking into account in this example the minimum and maximum MCS indexes only. The fair share is defined as the maximum capacity of the wireless link between each FAP and the FCR UAV, and assumed to be equal to the data rate of $MCS_i$ index over the number of FAPs sharing the medium. The minimum $SNR_i$ required for using $MCS_i$ imposes a minimum received power $P_{r_i}$ at $FAP_i$. Then, considering a transmission power $P_{t_i}$ for $FAP_i$, initially set to 0 dBm, EREP calculates the maximum transmission range $r_i$ for $FAP_i$, in order to achieve the minimum $SNR_i$, as presented in \cref{fig:imageA} for $FAP_1$ and $FAP_2$ (line 1 of Alg. 1); we assume $P_{t_i}$ is equal for all FAPs. In 3D space, $r_i$ represents the radius of a sphere centered at $FAP_i$. Next, EREP finds the volume that results from the intersection of the spheres centered at each FAP; the volume resulting from the intersection of the spheres centered at two FAPs is illustrated in \cref{fig:imageA} (lines 3, 4, and 8 of Alg. 1). The intersection between the spheres defines the volume inside which the FCR UAV can move without compromising the QoS. If no intersection is found, $P_{t_i}$ is successively increased by 1 dBm ($r_i$ increases) until an intersection between the spheres occurs (lines 5 and 6 of Alg. 1). 

Once an intersection volume is found, the altitude corresponding to the highest area inside the intersection volume is selected (line 11 of Alg. 1). A constant altitude for the trajectory to be completed by the FCR UAV is defined, since significant changes in the UAV altitude imply additional power consumption \cite{demir2019energy, Zorbas2016, Abeywickrama2018}. The centroid of that area is the position defined by the GWP algorithm, where all possible trajectories for the FCR UAV must pass through (line 12 of Alg. 1). The next step consists in defining the waypoints for the possible trajectories. EREP computes three possible trajectories and five waypoints that maximize the length of each trajectory, so that the FCR UAV can move as long as possible. Three different trajectories are calculated due to the different shapes of the area of intersection, which varies according to the distances between the FAPs and the number of FAPs. An example of the three trajectories computed is given in \cref{fig:set4images}. For the first trajectory (\cref{fig:imageB}), apart from the centroid, the waypoints are defined as the points in the intersection area with the highest and lowest values of $x$ in both extremes of the $y$-axis (line 13 of Alg. 1). For the second trajectory  (\cref{fig:imageC}), the waypoints are defined as the points in the intersection area with the highest and lowest values of $y$ in both extremes of the $x$-axis (line 13 of Alg. 1). For the third trajectory  (\cref{fig:imageD}), the waypoints are defined as the four extreme points in the area of intersection that have the same $x$ or $y$ coordinate as the centroid (line 13 of Alg. 1). The selected trajectory is the one that has the highest total sum of distances between successive waypoints (line 14 of Alg. 1), in order to maximize the time the FCR UAV is moving at the speed consuming the lowest power. 

\begin{table}
    \sisetup{detect-weight=true}
	\renewcommand{\arraystretch}{1.4}
	\def\arraystretch{1.8}
	\caption{Relation between SNR, data rate of the minimum and maximum IEEE 802.11ac MCS indexes, and fair share for 2 FAPs that use the same wireless channel~\cite{MCSIndexTable:online}. The same rationale is valid for other MCS indexes and different number of FAPs.}
	\label{tab:snr-fair-share}
	\centering
	\begin{tabular}{|c|c|c|}
		\hline
		\textbf{SNR (\SI{}{dB})} & \textbf{MCS data rate (\SI{}{Mbit/s})} & \textbf{Fair share (\SI{}{Mbit/s})} \\ 
		\hline \hline
		$11$ & $58.5$ & $\frac{58.5}{2} = 29$ \\ \hline
		$38$ & $780$ & $\frac{780}{2} = 390$ \\ \hline
	\end{tabular}
\end{table}

The trajectory is defined by 5 waypoints: the centroid (Pc) and the edges of the area (P1, P2, P3, and P4). The UAV starts in Pc and goes to P1. Afterward, it moves to P2 and then to P3, passing through Pc. Before returning to Pc, the UAV passes through P4. The UAV hovers for \SI{1}{\second} at each of the waypoints to invert the movement direction. The 1 second hovering is used as an approximation to the energy consumed during the change of direction. This was considered in EREP because, to the best of our knowledge, there is no model in the state of the art available to characterize the energy consumption for this action of the UAV. The 1 second hovering approximation considered by the EREP algorithm is represented in the optimization problem defined in~\cref{eq:optimization-problem} by a set of consecutive equal waypoints $Q_0 (t)$ during 1 second.

In EREP, the transmission range of the wireless link established between each FAP and the FCR UAV is represented by a sphere centered at each FAP. The volume resulting from the intersection of all spheres (cf. \cref{fig:imageA}) corresponds to the \emph{IntersectionPoints} variable in \cref{gwp}. \emph{DesiredAltitude} is the $z$ coordinate where the $xy$-plane with the largest area is located, considering the intersection volume between the spheres (\emph{IntersectionPoints} variable). This is the plane that allows to define the longest trajectory to be completed by the FCR UAV, thus contributing to minimize the energy consumption (lowest energy consumption when compared to hovering). The centroid corresponds to the position of that plane that minimizes the distance between the FCR UAV and all the FAPs. Finally, the \emph{maximumDistance} variable represents the Euclidean distance of the longest trajectory amongst the three trajectories (cf. \cref{fig:imageB}, \cref{fig:imageC}, and \cref{fig:imageD}) defined in that plane by EREP, considering as input the Cartesian coordinates of the waypoints to be covered by the FCR UAV for completing each corresponding candidate trajectory.

\begin{algorithm}
\caption{Energy-aware RElay Positioning Algorithm}\label{gwp}
\begin{algorithmic}[1]
\State Set target SNR values and transmission range for each FAP \Comment{Using the Free-Space Path Loss model}
\State IntersectionPoints=$\varnothing$ \Comment{Volume resulting from the intersection of all spheres (transmission ranges of all FAPs)}
\While {IntersectionPoints is empty}
\State Calculate Intersection \Comment{Determine the intersection between the spheres centered at the FAPs, if possible (cf. \cref{fig:imageA})}
\If{No intersection is found}
\State Increase $P_t$ for all FAPs by 1dBm \Comment{Increase transmission power of all FAPs, initially set to 0 dBm, by 1 dBm}
\Else
\State IntersectionPoints $\gets$ Intersection \Comment{Volume determined after the intersection of all spheres has been achieved}
\EndIf
\EndWhile\label{IntersectionWhile}
\State DesiredAltitude $\gets$ Altitude with more points \Comment{$xy$-plane enabling the longest trajectory to be completed by the FCR UAV}
\State Find the centroid \Comment{Position at $DesiredAltitude$ that minimizes the distance between the FCR UAV and all FAPs}
\State Define waypoints for the first, second and third trajectories \Comment{According to \cref{fig:imageB}, \cref{fig:imageC}, and \cref{fig:imageD}.}
\State Selected Trajectory $\gets$ maximumDistance(first, second, third) \Comment{Longest trajectory (Euclidean distance) of the three defined}
\end{algorithmic}
\end{algorithm}

\section{Performance Evaluation}\label{sec:PerformanceEvaluation}
This section presents the evaluation of the EREP algorithm considering both the FCR UAV energy consumption and the corresponding flying network performance. It includes the simulation setup, the simulation scenarios defined, the performance metrics considered, and the simulation results obtained.  

\subsection{Energy Consumption Evaluation}\label{sec:energy-consumption-evaluation}
To perform the evaluation of the EREP algorithm in terms of the FCR UAV energy consumption, we developed a custom simulator in \emph{Python}, named UAVPowerSim, implementing the EREP algorithm and the power consumption model presented in \cref{eq:power-consumption}. The simulator is publicly available \cite{UAVPowerSimulator}. The simulator was used to evaluate the power consumption when the FCR UAV moves along the trajectory defined by EREP against the baseline -- the FCR UAV hovering at Pc, which is defined by the GWP algorithm \cite{coelho2019traffic}. The physical attributes of the UAV and the environment constants considered were the ones used by Zeng et al. \cite{zeng2019energy}. Those values correspond to the UAV weight, $W$ = \SI{20}{\newton}, rotor radius, $R$ = \SI{0.4}{\meter}, blade angular velocity, $\Omega$ = \SI{300}{\radian/\second}, incremental correction factor to induced power, $k$ = 0.1, profile drag coefficient, $\delta$ = 0.012, \textcolor{blue}{air density,} $\rho$ = \SI{1.225}{\kilogram/\cubic\metre}, \textcolor{blue}{rotor disc area,} $A=\pi \cdot R^2$ = \SI{0.503}{\square\meter}, \textcolor{blue}{tip speed of the rotor blade,} $U_{tip}\triangleq \Omega R$ = \SI{120}{\meter/\second}, \textcolor{blue}{fuselage drag ratio,} $d_0$ = 0.6, \textcolor{blue}{mean rotor induced velocity in hovering
state,} $v_0= \sqrt{\frac{W}{2\rho A}}$ = 4.03, \textcolor{blue}{rotor solidity,} $s$ = 0.05, \textcolor{blue}{blade profile power in hovering state, } $P_b\triangleq\frac{\delta}{8}\rho sA\Omega^3R^3$ $\approx$ 79.86, \textcolor{blue}{and induced power in hovering state,} $P_{ind}\triangleq(1+k)\frac{W^{3/2}}{\sqrt{2\rho A}}$ $\approx$ 88.63. The main simulation parameters are presented in~\cref{tab:simulation-parameters}. For these values, the speed that minimizes the power consumption is $V$ $\approx$ \SI{10.2}{\meter/\second}.  

\begin{figure}[!ht]
	\centering
	\includegraphics[width=0.8\linewidth]{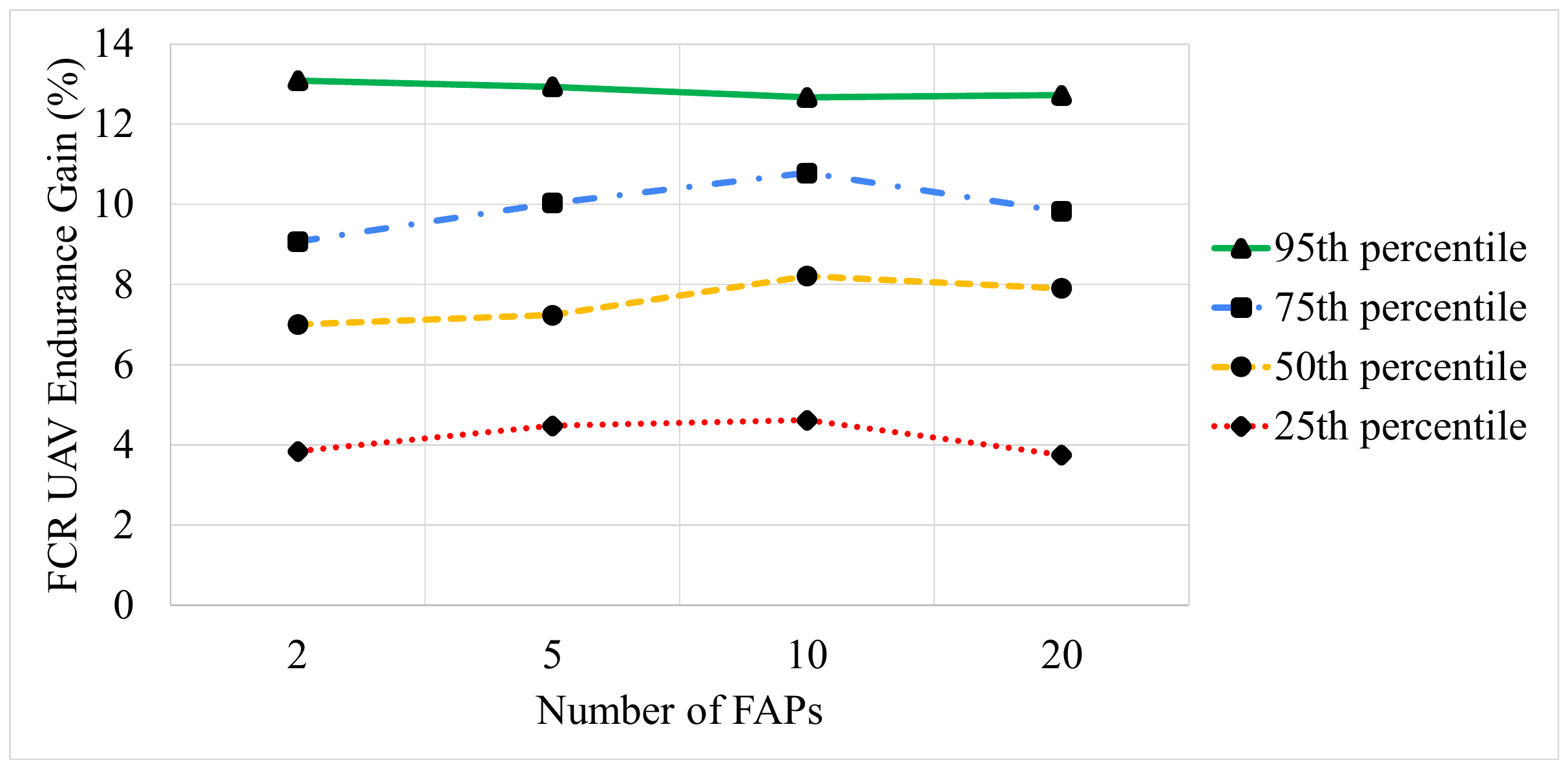}
	\caption{25\textsuperscript{th}, 50\textsuperscript{th}, 75\textsuperscript{th}, and 95\textsuperscript{th} percentiles of the FCR UAV endurance gains (\%) achieved by the EREP algorithm, considering 160 random networking scenarios for each number of FAPs between 2 and 20. The baseline (0\% gain) considers the FCR UAV hovering at the position defined by the GWP algorithm \cite{coelho2019traffic}.}
	\label{fig:results}
\end{figure}

The evaluation of the EREP algorithm was performed under random networking scenarios, considering 2, 5, 10, and 20 FAPs. For that purpose, a set of 160 networking scenarios for each number of FAPs was generated using BonnMotion \cite{aschenbruck2010bonnmotion}, a mobility scenario generation tool. In the simulations for the same number of FAPs using UAVPowerSim, all the FAPs were generating the same amount of traffic. The results are shown in \cref{fig:results}. Since we considered a set of 160 random networking scenarios for each number of FAPs, the gains regarding the FCR UAV endurance ($T$) are presented by means of the 25\textsuperscript{th}, 50\textsuperscript{th}, 75\textsuperscript{th}, and 95\textsuperscript{th} percentiles, considering as baseline the endurance achieved when using the GWP algorithm. From the results it is possible to observe that the average gains oscillate between approximately 7\% and 8\%, while the 95\textsuperscript{th} percentile gains go up to 13\%. Overall, the gains are maximized when the distance between the FAPs is minimized, since the volume resulting from the intersection of the spheres centered at the FAPs is maximized. Considering that the performance evaluation carried out takes into account random positions for the FAPs, the maximum gains achieved by EREP can be even greater if the positions of the FAPs are defined to maximize the volume of intersection between the corresponding spheres. On the other hand, the FCR UAV's endurance gains basically do not depend on the number of FAPs; this is mainly due to the volume of intersection between the FAPs, which has roughly the same value regardless of the number of FAPs, since the intersection is found when the minimum value of transmission power $P_{t_i}$ for it to exist is achieved. A gain of 13\% means that for a drone with a 2-hour endurance, it will keep flying for 16 minutes more when compared against the baseline, which considers the FCR UAV hovering at the position defined by the GWP algorithm. The evaluation of EREP against the baseline was carried out under the same exact conditions and assumes a negligible wind speed. The analysis of the wind effect on the performance achieved by EREP is left for future work.

\begin{table}
	\caption{Main simulation parameters.}
	\label{tab:simulation-parameters}
	\centering
	\begin{tabular}{|c|c|}
		\hline
		\textbf{Parameter} & \textbf{Value} \\ 
		\hline \hline
		IEEE 802.11 standard & IEEE 802.11ac \\ \hline
		Guard Interval (GI) & \SI{800}{\nano\second} \\ \hline
		Channel bandwidth & \SI{160}{\mega\hertz} \\ \hline
		Spatial streams & 1 \\ \hline
		Maximum physical data rate & \SI{780}{Mbit/s} \\ \hline
		Carrier frequency ($f$) & \SI{5180}{\mega\hertz} \\ \hline
		Speed of light in vacuum ($c$) & \SI{3E8}{\meter/\second} \\ \hline
		Transmission power ($P_t$) & \SI{20}{dBm} \\ \hline
		Noise power ($N_0$) & \SI{-85}{dBm} \\ \hline
		UAV weight ($W$) & \SI{20}{\newton} \\ \hline
		Rotor radius ($R$) & \SI{0.4}{\meter} \\ \hline
		Blade angular velocity ($\Omega$) & \SI{300}{\radian/\second} \\ \hline
		Incremental correction factor to inducer power ($k$) & 0.1 \\ \hline
		Profile drag coefficient ($\delta$) & 0.012 \\ \hline
		Air density ($\rho$) & \SI{1.225}{\kilogram/\cubic\metre} \\ \hline
		Rotor disc area ($A$) & \SI{0.503}{\square\meter} \\ \hline
		Tip speed of the rotor blade ($U_{tip}$) & \SI{120}{\meter/\second} \\ \hline
		Fuselage drag ratio ($d_0$) & 0.6 \\ \hline
		Mean rotor induced velocity in hovering state ($v_0$) & 4.03 \\ \hline
		Rotor solidity ($s$) & 0.05 \\ \hline
		Blade profile power in hovering state ($P_b$) & 79.86 \\ \hline
		Induced power in hovering state ($P_{ind}$) & 88.63 \\ \hline
	\end{tabular}
\end{table}

The problem addressed by the EREP algorithm is an evolution of the problem addressed by the GWP algorithm~\cite{coelho2019traffic}. While the GWP algorithm aims at improving the overall flying network performance, considering the minimum required transmission power $P_t$ for the Network Interface Cards carried on board the UAVs to reduce interference, the EREP algorithm aims at minimizing the power consumed by the FCR UAV for propulsion. GWP places the FCR UAV in hovering state at a fixed position, which is not the most energy-efficient UAV state~\cite{zeng2019energy}. The considerable number of networking scenarios used in the performance evaluation provides statistical relevance to the results presented.

\subsection{Network Performance Evaluation}
In order to assess the flying network performance when using the EREP algorithm, the ns-3 simulator~\cite{ns3Simulator} was used. In ns-3, each simulated UAV was carrying a Network Interface Card configured with \SI{20}{dBm} transmission power and using the IEEE 802.11ac standard operating in Ad Hoc mode and in channel 50, which enables \SI{160}{\mega\hertz} channel bandwidth. The wireless links established between the FCR UAV and the FAPs were using a single spatial stream and \SI{800}{\nano\second} GI. UDP Poisson traffic using data packets of 1400 bytes was considered. The data rate was automatically defined by the \emph{MinstrelHtWifiManager} mechanism and the transmission queues were controlled by the Controlled Delay (CoDel) algorithm~\cite{queue}, using the default parameters in ns-3~\cite{ns3Codel}, in order to mitigate the bufferbloat problem~\cite{gettys2011}. A summary of network parameters is included in~\cref{tab:simulation-parameters}. 

A set of networking scenarios consisting of 2, 5, and 10 FAPs at different distances from each other were considered. They represent extreme scenarios in terms of the distance between the FAPs, aiming at evaluating the flying network performance achieved when the FAPs are positioned close to each other and away from each other. The random Cartesian coordinates of the FAPs for each networking scenario are presented in~\cref{tab:network-performance-results}. The traffic demand of each FAP was defined based on an estimation for the maximum capacity achievable in the wireless channel, which we assume to be equal to approximately 65\% of the data rate that characterizes the highest Modulation and Coding Scheme (MCS) index for the network configuration used ($65\% \times \SI{780}{Mbit/s} \approx \SI{500}{Mbit/s}$) over the number of FAPs sharing the medium, in order to take into account the MAC inefficiency and consider a margin with respect to the theoretical maximum channel capacity (\SI{780}{Mbit/s}).

The network performance evaluation carried out took into account two metrics: 
\begin{itemize}
    \item \textbf{Aggregate throughput:} the number of bits received per second by the FCR UAV.
    \item \textbf{Delay:} the time taken for each data packet to be delivered to the sink application of the FCR UAV, since the instant the packet was sent by the source application of each FAP, including queueing, transmission, and propagation delays. It was measured considering the packets collected every \SI{10}{\milli\second}.
\end{itemize}
The results for each scenario, considering all values collected in 20 simulation runs, are represented by means of the Cumulative Distribution Function (CDF) for the delay and by the Complementary CDF (CCDF) for the aggregate throughput. The CDF $F(x)$ gives the percentage of samples for which the delay is less than or equal to $x$, while the CCDF $F'(x)$ gives the percentage of samples for which the throughput is greater than $x$. For analysis purposes, the values obtained for the 90\textsuperscript{th} percentile are considered.

\begin{table}[!ht]
	\renewcommand{\arraystretch}{1.4}
	\def\arraystretch{1.8}
	\caption{Performance results of EREP compared with the GWP algorithm, under networking scenarios composed of 2, 5, and 10 FAPs positioned at different distances between them, considering the 90\textsuperscript{th} percentile.}
	\label{tab:network-performance-results}
	\centering
	\begin{tabular}{|c|M{0.14\columnwidth}|M{0.14\columnwidth}|M{0.14\columnwidth}|M{0.14\columnwidth}|}
		\hline
		\textbf{Networking scenario} & \textbf{FAPs' Cartesian coordinates} & \textbf{Throughput degradation (\%)} & \textbf{Delay degradation (\%)} & \textbf{UAV endurance gain (\%)} \\ 
		\hline \hline
		2 FAPs close to each other & (0, 0, 10), (1, 0, 10) & 7 & 20 & 26 \\ \hline
		2 FAPs away from each other & (0, 0, 10), (58, 0, 10) & 2 & 0 & 7 \\ \hline
		5 FAPs close to each other & (19, 40, 12), (1, 0, 10), (7, 17, 17), (9, 16, 7), (10, 36, 13) & 0 & 5 & 19 \\ \hline
		5 FAPs away from each other & (30, 32, 2), (3, 45, 0), (43, 4, 6), (23, 3, 7), (2, 16, 15) & 1 & 2 & 4 \\ \hline
		10 FAPs close to each other & (20, 25, 18), (9, 20, 17), (20, 13, 5), (24, 35, 13), (20, 40, 7), (35, 42, 12), (41, 30, 15), (40, 25, 1), (14, 43, 17), (29, 19, 13) & 2 & 5 & 20 \\ \hline
		10 FAPs away from each other & (41, 48, 14), (44, 3, 15), (16, 4, 3), (11, 9, 2), (40, 36, 5), (24, 35, 15), (29, 40, 8), (46, 32, 14), (3, 11, 16), (25, 27, 6) & 5 & 3 & 5 \\ \hline
	\end{tabular}
\end{table}

\begin{figure}[!ht]
    \begin{minipage}[b]{0.49\textwidth}
    \centering
	    \subfloat[Throughput CCDF.]{
	        \includegraphics[width=0.70\columnwidth]{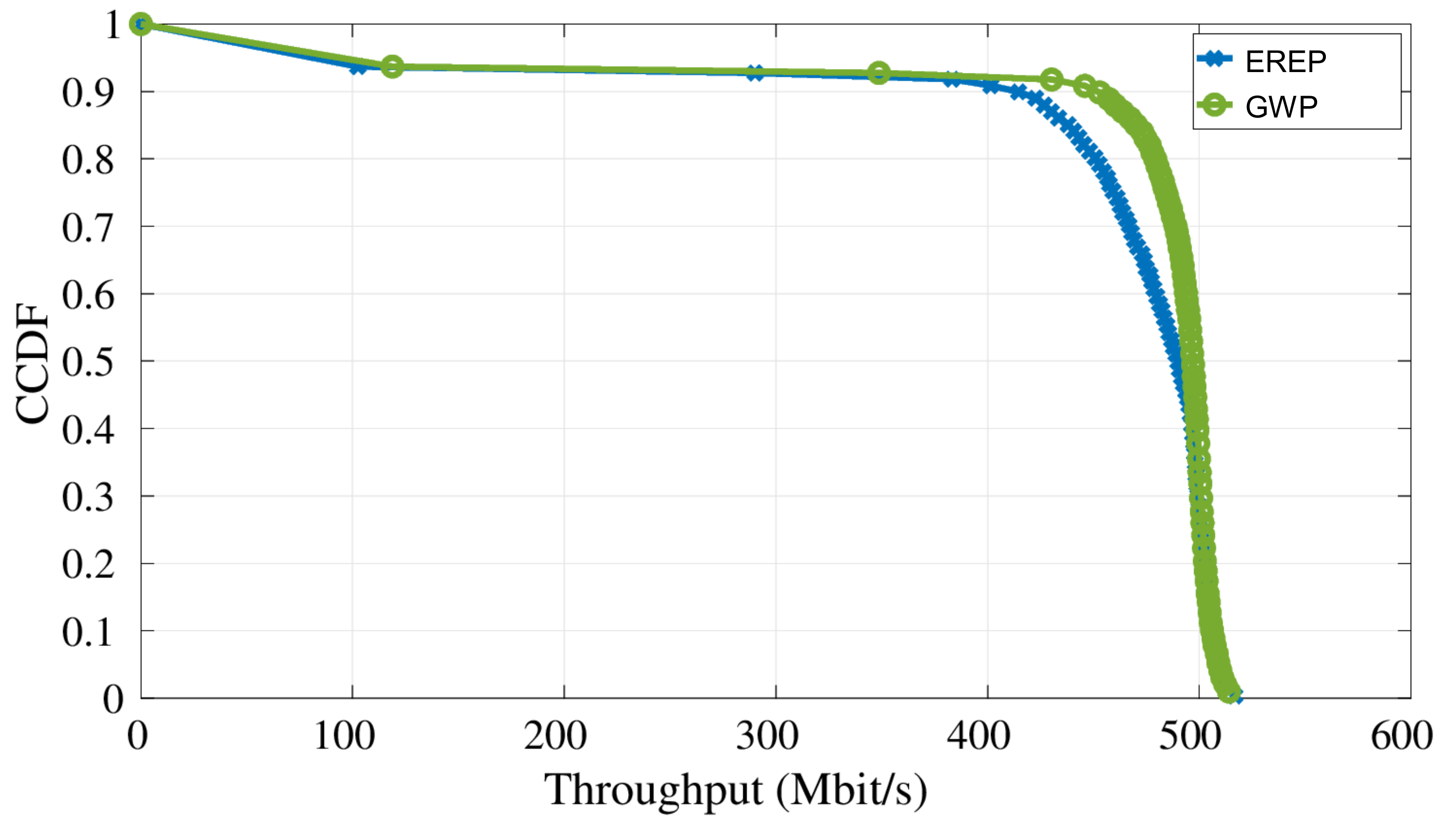}
		    \label{fig:throughput-2-faps-close-each-other}}
		    \hfill
	    \subfloat[Delay CDF.]{
		    \includegraphics[width=0.70\columnwidth]{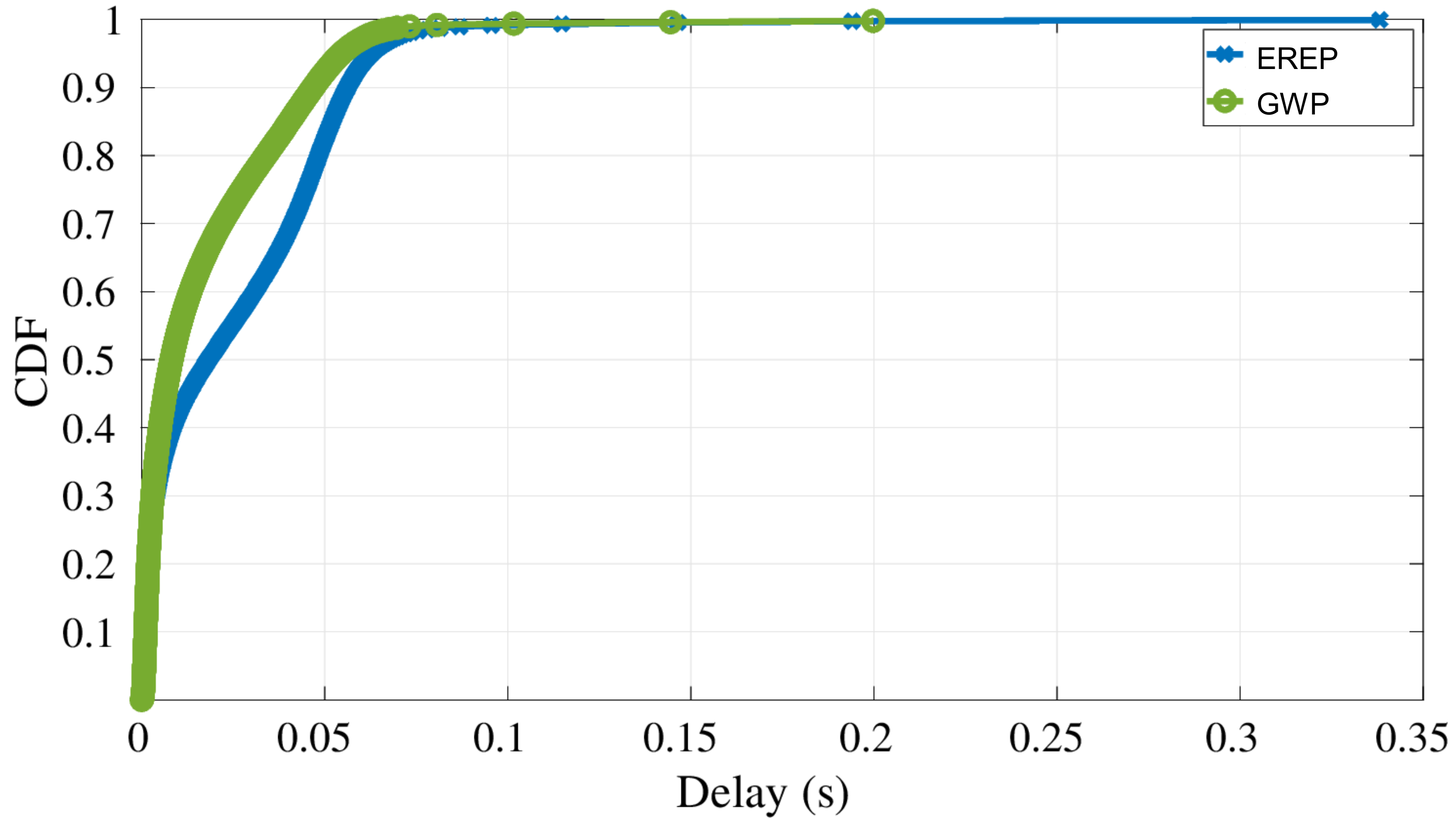}
		    \label{fig:delay-2-faps-close-each-other}}
	    \caption{2 FAPs close to each other.}
	\label{fig:2-faps-close-each-other}
    \end{minipage}
    \hfill
    \begin{minipage}[b]{0.49\textwidth}
    \centering
	    \subfloat[Throughput CCDF.]{
		    \includegraphics[width=0.70\columnwidth]{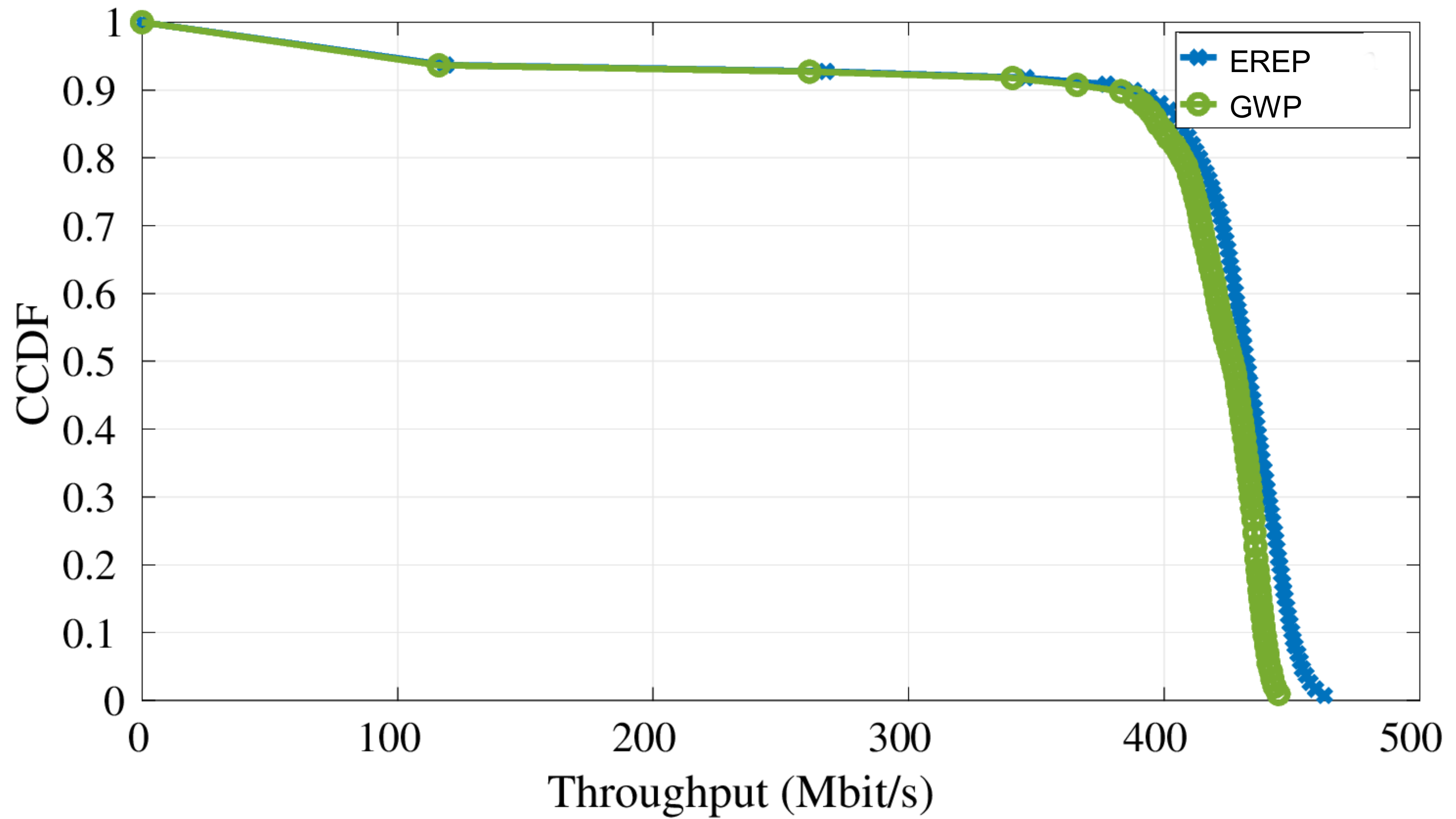}
		    \label{fig:throughput-2-faps-away-from-each-other}}
		    \hfill
	    \subfloat[Delay CDF.]{
		    \includegraphics[width=0.70\columnwidth]{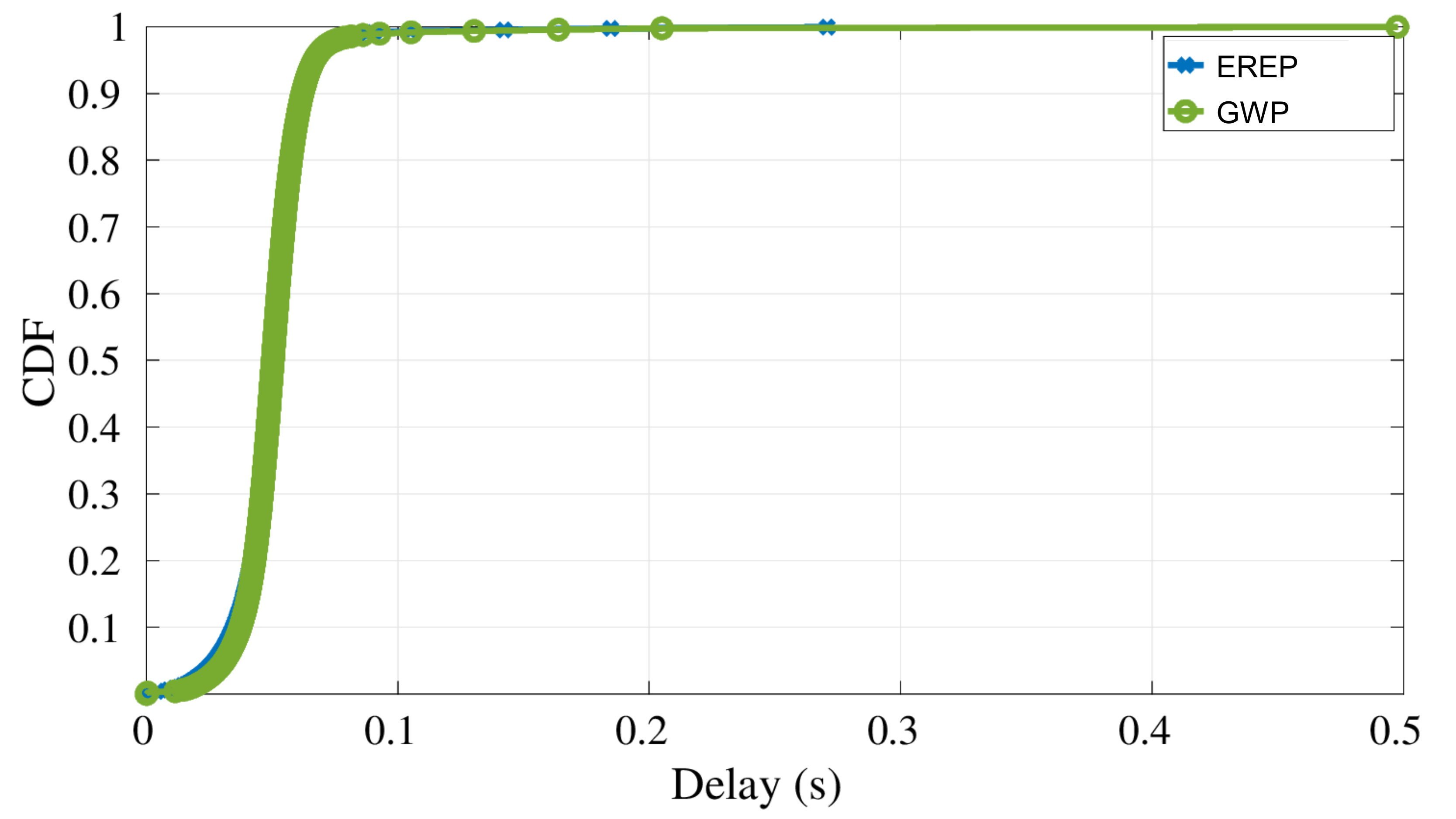}
		    \label{fig:delay-2-faps-away-from-each-other}}
	    \caption{2 FAPs away from each other.}
	\label{fig:2-faps-away-from-each-other}
	\end{minipage}
\end{figure}

\begin{figure}[!ht]
    \begin{minipage}[b]{0.49\textwidth}
    \centering
	    \subfloat[Throughput CCDF.]{
	        \includegraphics[width=0.70\columnwidth]{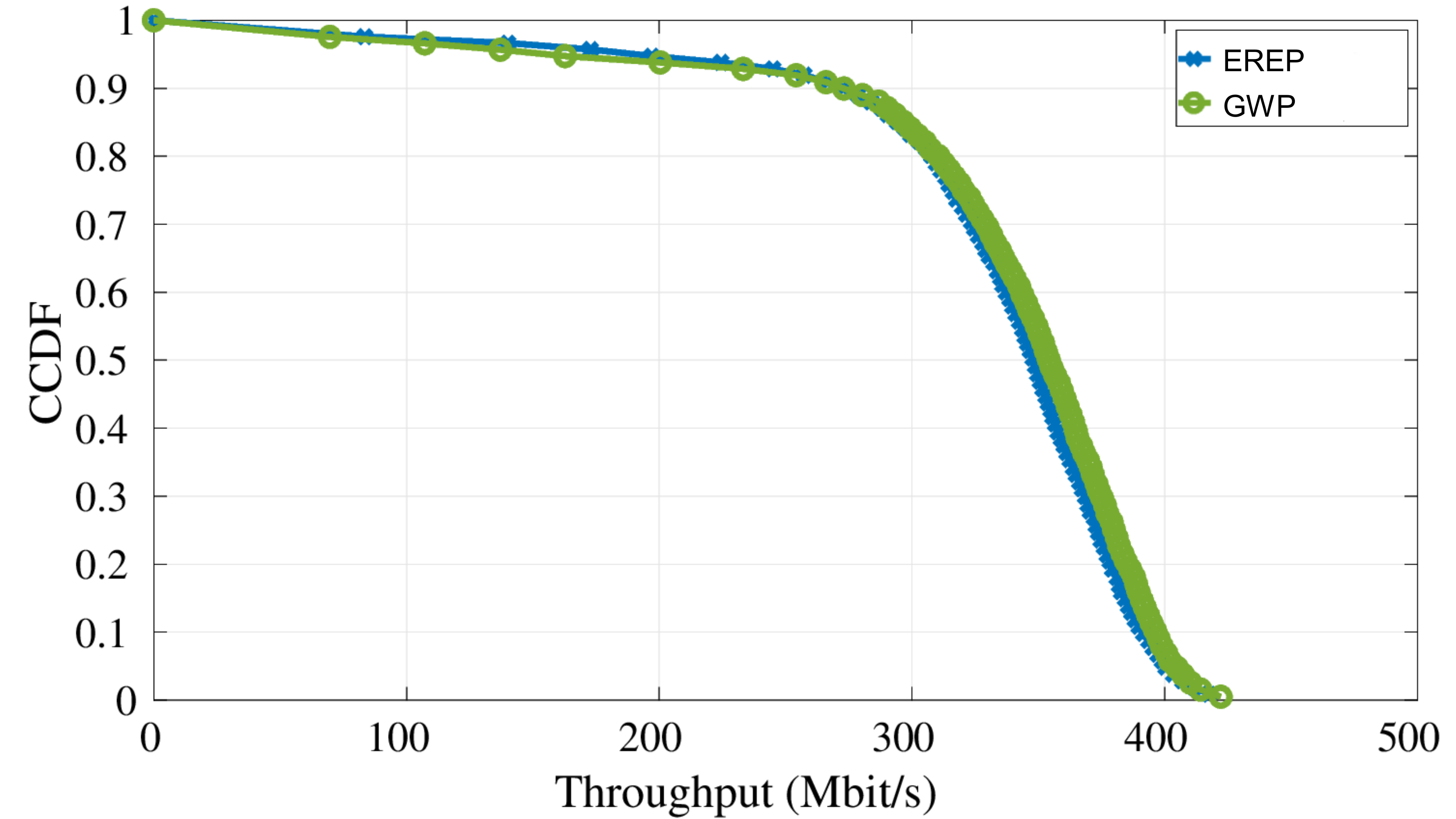}
		    \label{fig:throughput-5-faps-close-each-other}}
		    \hfill
	    \subfloat[Delay CDF.]{
		    \includegraphics[width=0.70\columnwidth]{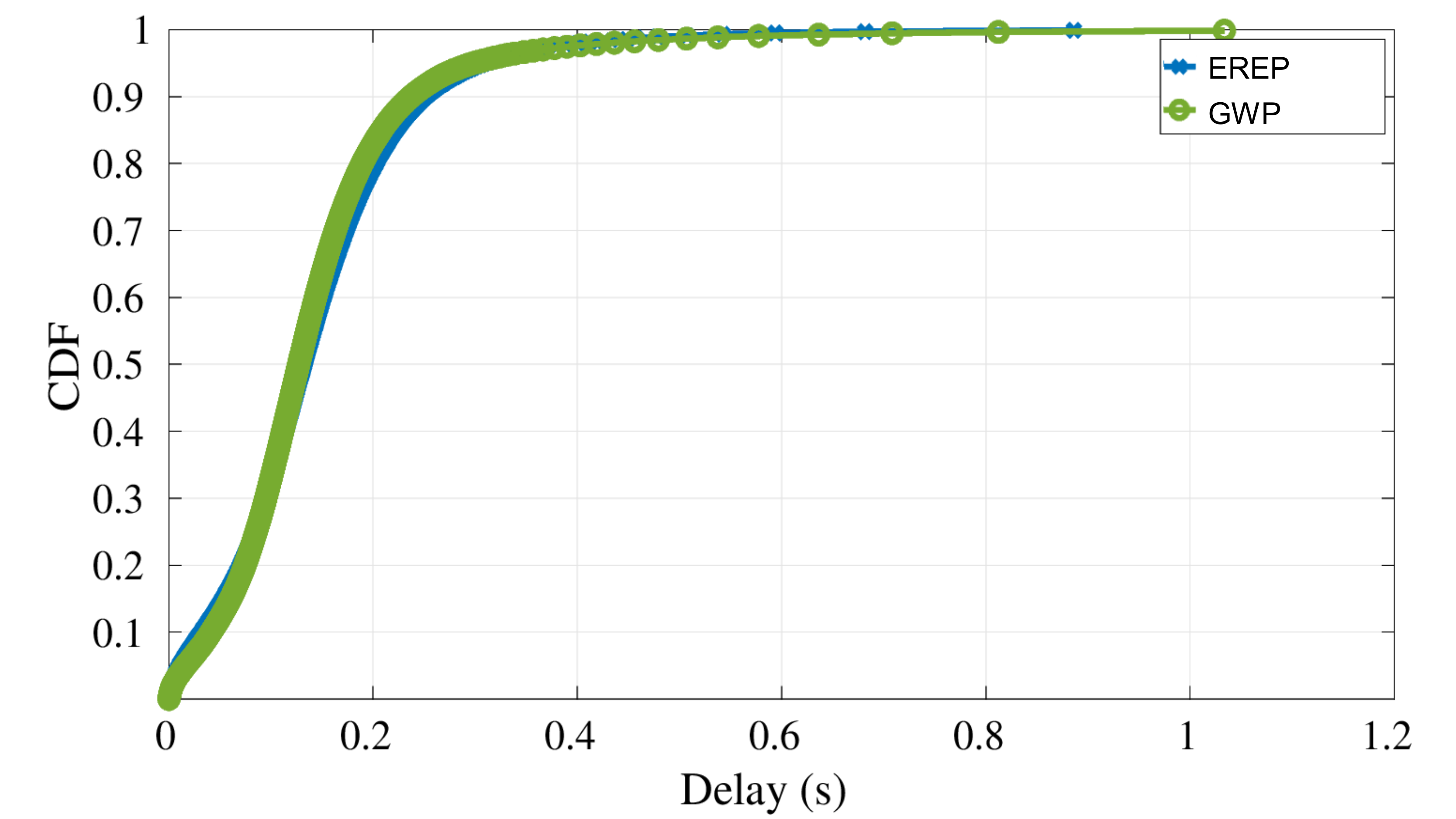}
		    \label{fig:delay-5-faps-close-each-other}}
	    \caption{5 FAPs close to each other.}
	\label{fig:5-faps-close-each-other}
    \end{minipage}
    \hfill
    \begin{minipage}[b]{0.49\textwidth}
    \centering
	    \subfloat[Throughput CCDF.]{
		    \includegraphics[width=0.70\columnwidth]{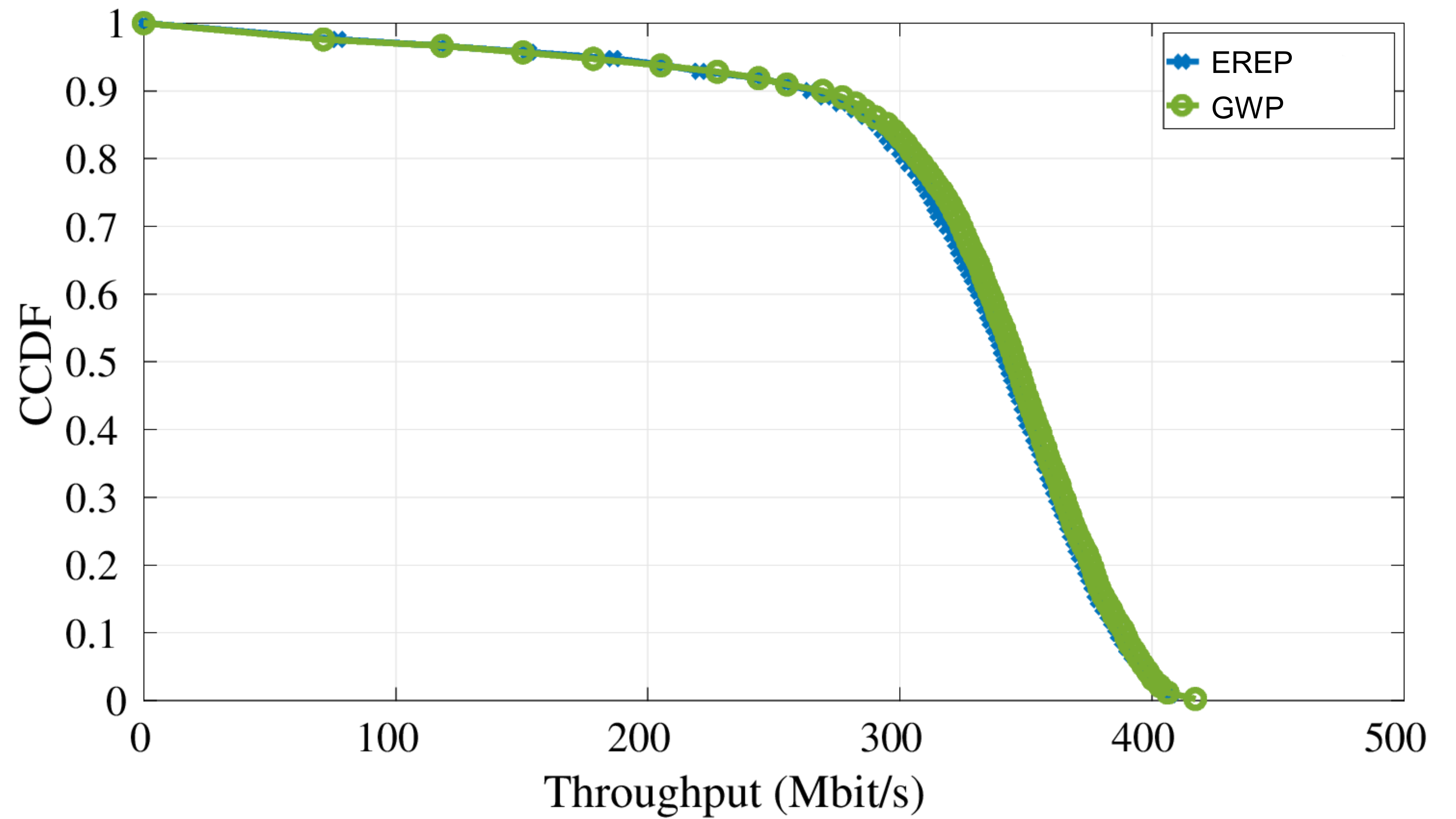}
		    \label{fig:throughput-5-faps-away-from-each-other}}
		    \hfill
	    \subfloat[Delay CDF.]{
		    \includegraphics[width=0.70\columnwidth]{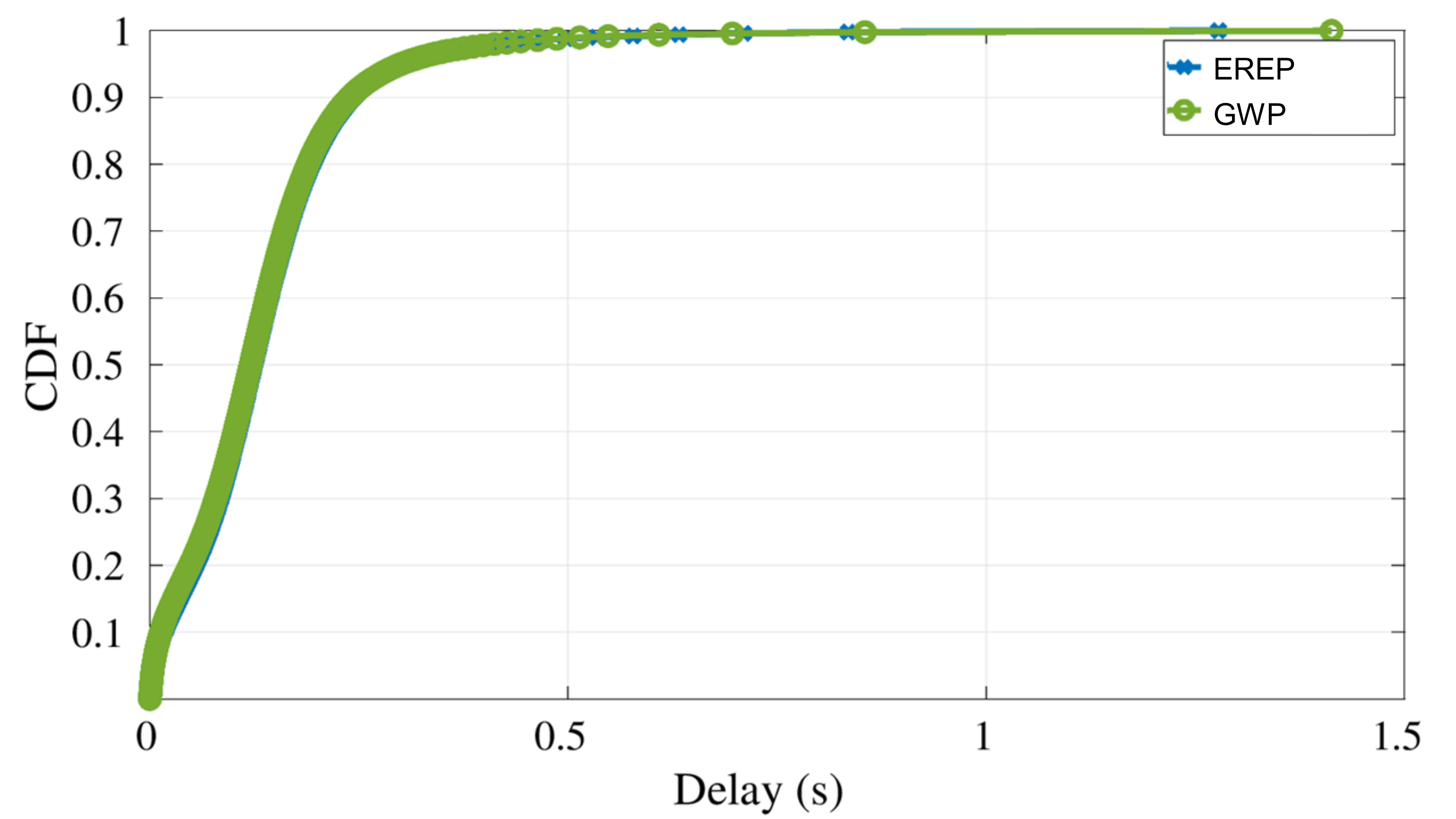}
		    \label{fig:delay-5-faps-away-from-each-other}}
	    \caption{5 FAPs away from each other.}
	\label{fig:5-faps-away-from-each-other}
	\end{minipage}
\end{figure}

\begin{figure}[!ht]
    \begin{minipage}[b]{0.49\textwidth}
	    \centering
	    \subfloat[Throughput CCDF.]{
	        \includegraphics[width=0.70\columnwidth]{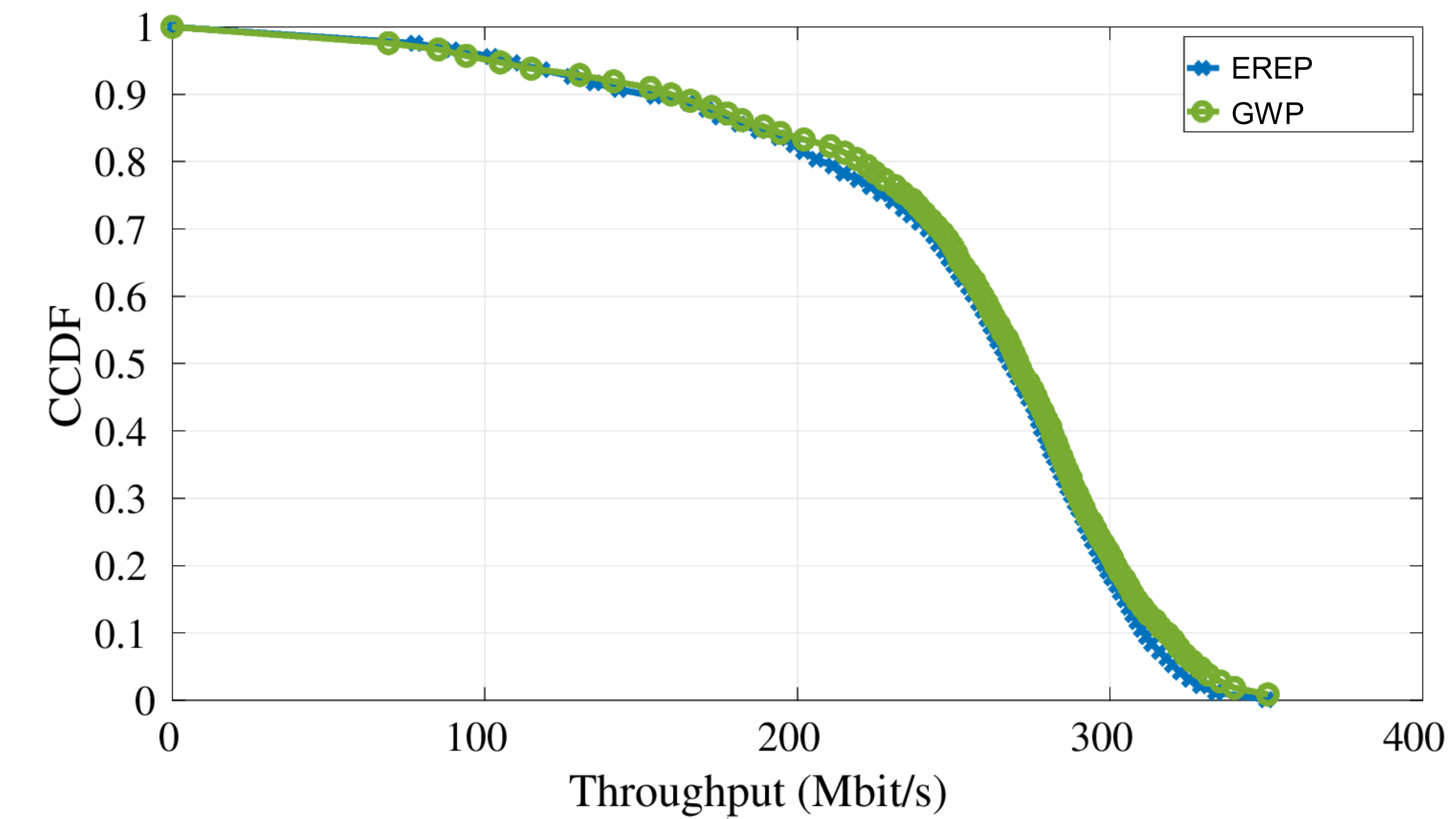}
		    \label{fig:throughput-10-faps-close-each-other}}
		    \hfill
	    \subfloat[Delay CDF.]{
		    \includegraphics[width=0.70\columnwidth]{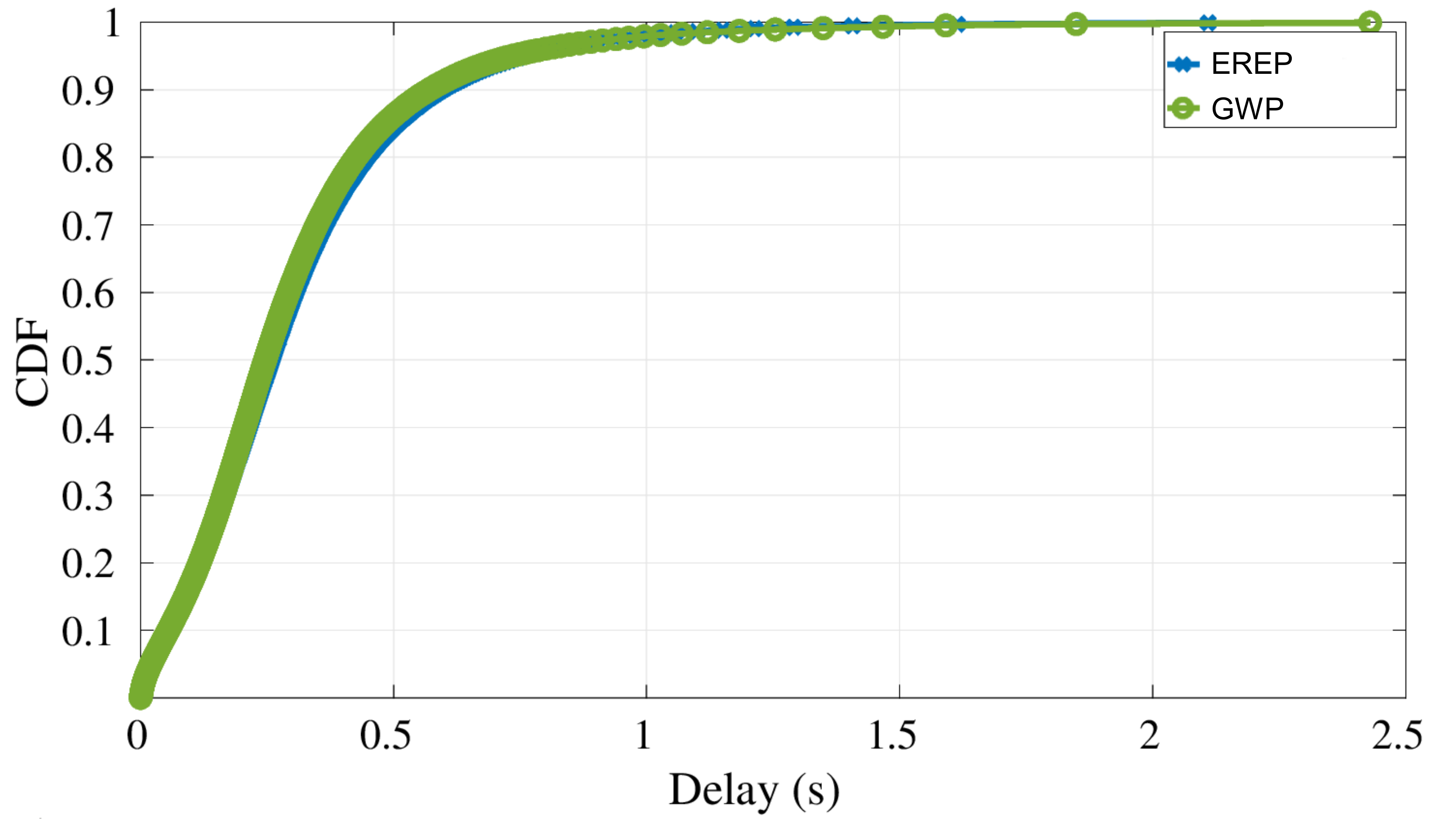}
		    \label{fig:delay-10-faps-close-each-other}}
	    \caption{10 FAPs close to each other.}
	\label{fig:10-faps-close-each-other}
    \end{minipage}
    \hfill
    \begin{minipage}[b]{0.49\textwidth}
	    \centering
	    \subfloat[Throughput CCDF.]{
		    \includegraphics[width=0.70\columnwidth]{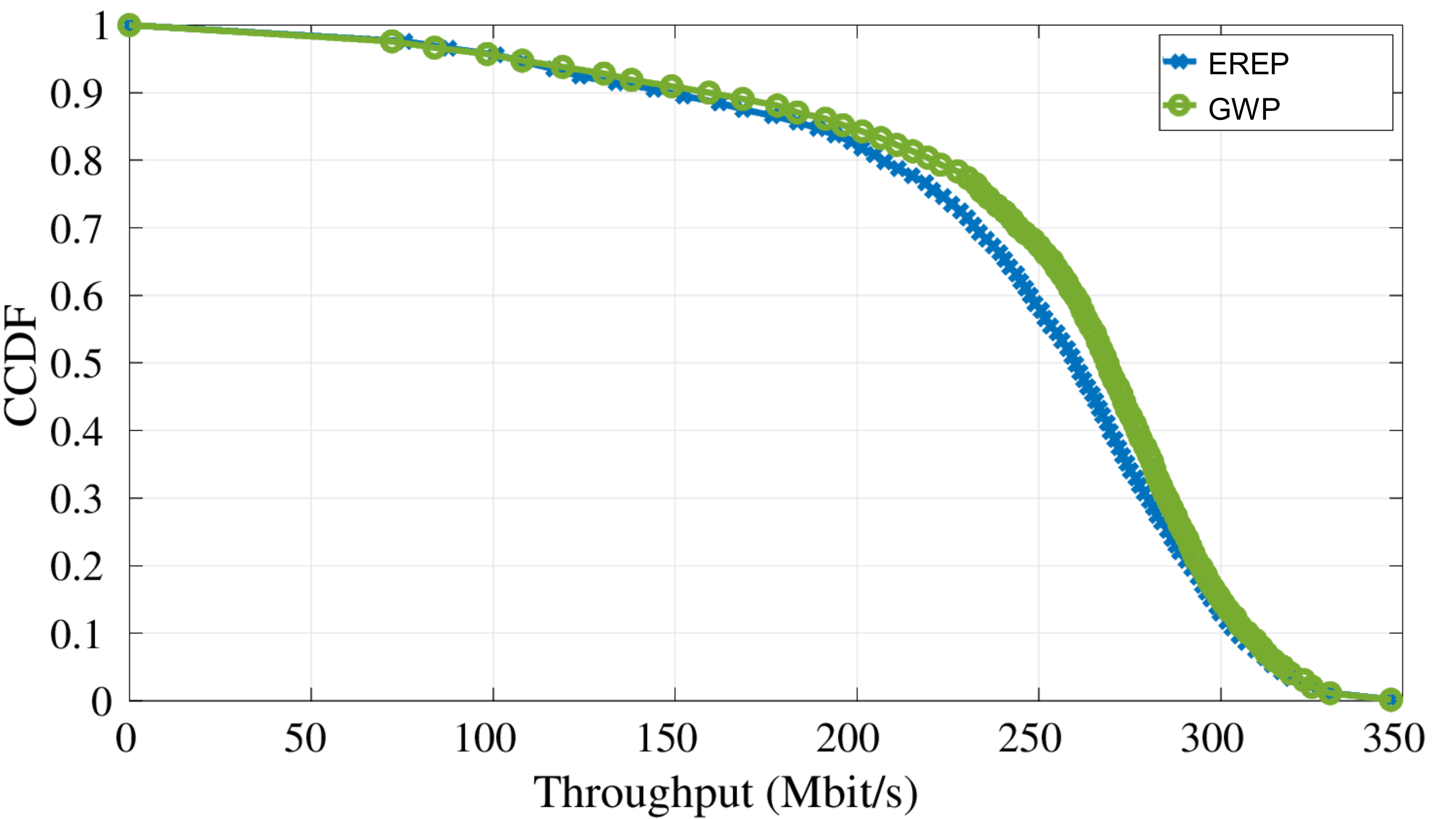}
		    \label{fig:throughput-10-faps-away-from-each-other}}
		    \hfill
	    \subfloat[Delay CDF.]{
		    \includegraphics[width=0.70\columnwidth]{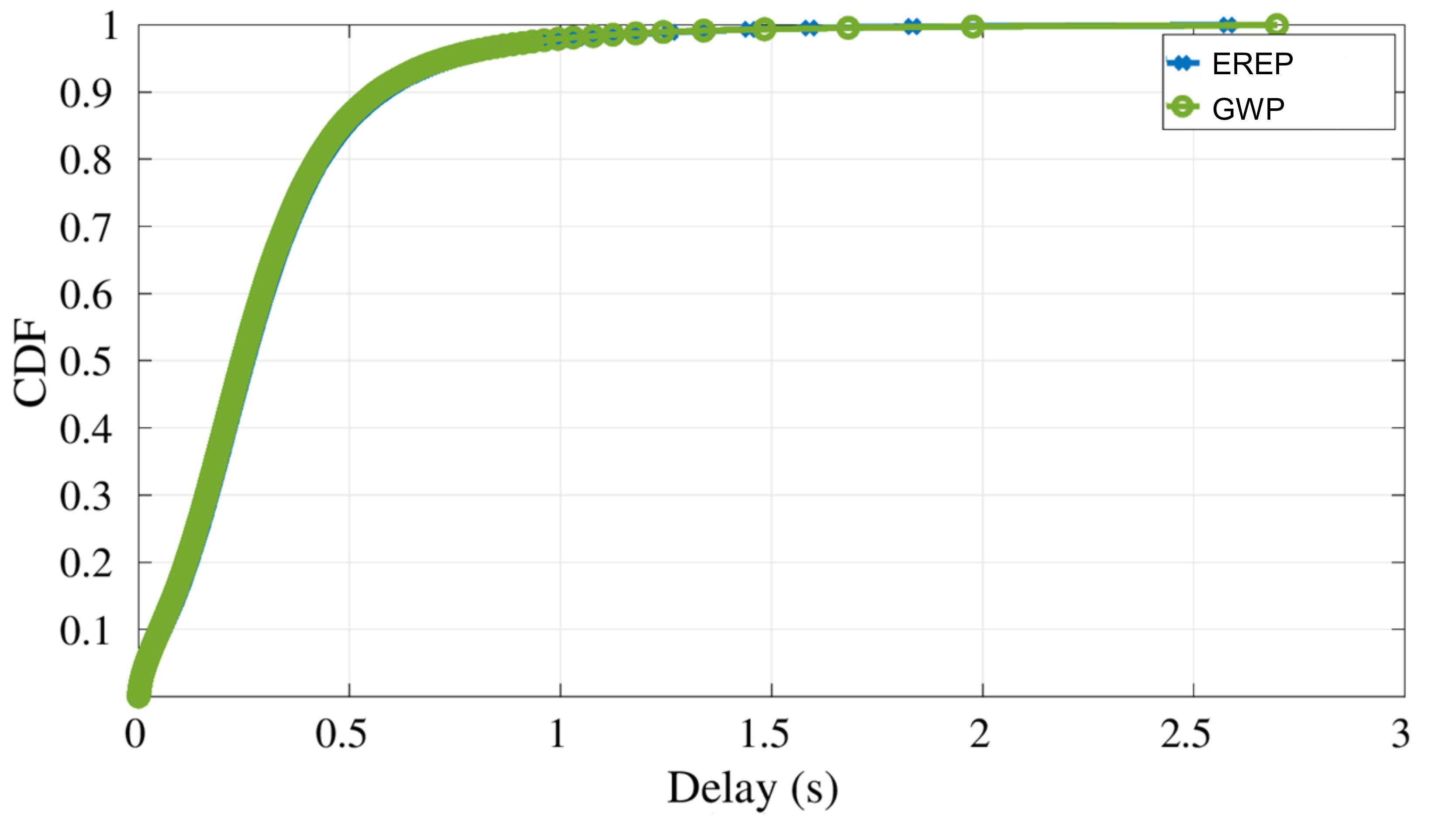}
		    \label{fig:delay-10-faps-away-from-each-other}}
	    \caption{10 FAPs away from each other.}
	\label{fig:10-faps-away-from-each-other}
	\end{minipage}
\end{figure}

The network performance results when considering 2, 5, and 10 FAPs positioned close to each other are depicted respectively in~\cref{fig:2-faps-close-each-other}, \cref{fig:5-faps-close-each-other}, and \cref{fig:10-faps-close-each-other}, while the results for the FAPs positioned away from each other are presented in~\cref{fig:2-faps-away-from-each-other}, \cref{fig:5-faps-away-from-each-other}, and \cref{fig:10-faps-away-from-each-other}. 
The networking scenario composed of 2 FAPs positioned close to each other (cf. \cref{fig:2-faps-close-each-other}) is the worst case scenario. In fact, for this networking scenario the volume resulting from the intersection of the spheres centered at the FAPs is large, since it is constrained by two FAPs only. This makes the distance allowed for the trajectory to be completed by the FCR UAV to be long, which leads to higher SNR degradation for the wireless links established between the FCR UAV and the FAPs, lower network capacity available, due to lower MCS indexes selected by \emph{MinstrelHtWifiManager}, and packets being held in the transmission queues longer as the distance increases. This justifies the slightly lower throughput (7\%) and higher delay (20\%) measured when using EREP. The performance is improved when the number of FAPs composing the networking scenario increases (cf. \cref{fig:5-faps-close-each-other} and \cref{fig:10-faps-close-each-other}), since the distance allowed for the trajectory to be completed by the FCR UAV decreases. Taking into account the results for 5 and 10 FAPs positioned close to each other (cf. \cref{fig:5-faps-close-each-other} and \cref{fig:10-faps-close-each-other} respectively), we observe that EREP does not compromise in practice the QoS offered when compared with GWP, which is our baseline. In fact, the results show the aggregate throughput is only reduced up to 2\%, while the delay is increased up to 5\%, which can be considered negligible differences.

When the FAPs are positioned away from each other (cf. \cref{fig:2-faps-away-from-each-other}, \cref{fig:5-faps-away-from-each-other}, and \cref{fig:10-faps-away-from-each-other}), the network performance is less affected when compared with the scenarios in which the FAPs are positioned close to each other. This is clearly shown for the networking scenarios composed of 2 FAPs only (\cref{fig:2-faps-close-each-other} vs. \cref{fig:2-faps-away-from-each-other}). Considering all the networking scenarios where the FAPs are positioned away from each other, we observe a throughput reduction up to 5\% and an increase in the delay up to 3\% when using EREP, taking into account the baseline results obtained with the GWP algorithm. Again, these are negligible differences that show EREP does not compromise in practice the QoS offered by the flying network, while reducing the energy consumed by the FCR UAV for propulsion.  

A summary of the obtained results, considering the different networking scenarios employed, is presented in~\cref{tab:network-performance-results}. In complement to the analysis presented in~\cref{sec:energy-consumption-evaluation}, we can conclude that, for the same number of FAPs, the FCR UAV endurance gains increase as the distance between the FAPs decreases; this is due to the increased volume of intersection between the spheres centered at the FAPs.

\subsection{Discussion\label{sec:discussion}}
In general, the gain in the UAV endurance achieved by EREP depends on the volume of intersection between the spheres centered at each FAP. For the same distance and number of FAPs, the gain will be higher if the transmission power $P_t$ is further increased to augment the volume of intersection on purpose, given the fact that the increase in a few \SI{}{dBm} in $P_t$ will be negligible when compared to the power consumption $P_0$ of the UAV. The higher the volume the higher the gain. However, it is important to note that EREP will always increase the UAV endurance, without compromising network performance.

In this article, the EREP algorithm is used only for the FCR UAV because it represents a single point of failure in the flying network. If the FCR UAV becomes unavailable due to energy shortage, the flying network will not be able to connect to the Internet, since the FCR UAV is the only communications node in charge of forwarding the traffic to/from the Internet. Some flexibility is inherent to the access network. Some FAPs can be removed for power charging during the network operation, and the transmission power and positioning of the remaining FAPs can be temporarily modified, in order to ensure the always-on coverage of the ground users, even if the network performance can be temporarily compromised. The usage of the EREP algorithm for the FAPs positioning brings additional challenges, since the movement of the FAPs must be performed so that the traffic demand of the ground users is met, while simultaneously ensuring the backhaul links established with the FCR UAV provide high enough capacity for accommodating the traffic demand. In addition, the collision avoidance problem should be considered, since multiple FAPs would be moving simultaneously. The improvement of EREP for multiple UAVs is left for future work. The current version of the EREP algorithm can also be used to define the positioning of the FCR UAV in a networking scenario composed of fixed terrestrial Wi-Fi Access Points or cellular Base Stations.

\textcolor{blue}{The EREP algorithm improves the FCR UAV endurance by defining a trajectory at the speed that consumes minimum power. At the same time, this trajectory aims at ensuring that the wireless link established between each FAP and the FCR UAV has a minimum SNR that induces the selection of a target MCS index by the auto rate mechanism used. The target MCS indexes are characterized by a data rate higher than or equal to the traffic offered by each FAP, in order to maximize the number of bits per second received in the FCR UAV. In its current version, the EREP algorithm allows to maximize the endurance of the FCR UAV. However,} the SNR experienced by the UAVs in real-world flying networks may present some deviation with respect to the theoretical values, leading to QoS degradation \textcolor{blue}{in some scenarios}, since EREP considers the Free-Space Path Loss model to calculate the minimum theoretical SNR values required for selecting the target MCS indexes. In order to overcome this challenge, the use of an SNR margin with respect to the theoretical values computed by means of the Free-Space Path Loss model may be considered. The use of the SNR margin will result in a smaller intersection volume between the spheres (transmission ranges) centered at the FAPs, leading to a shorter trajectory for the FCR UAV and avoiding network performance degradation due to the SNR degradation that can occur in some scenarios\textcolor{blue}{, at the expense of a lower FCR UAV endurance gain when compared with the current version of EREP}. This aspect is left for future work.

\section{Conclusions}\label{sec:Conclusions}
This article proposed an energy-aware relay positioning algorithm (EREP) for an FCR UAV, with the goal of minimizing the FCR UAV power consumption for propulsion. The EREP algorithm defines a trajectory to be completed by the FCR UAV, which is accomplished at the speed that consumes minimum power. The obtained simulation results show that the proposed algorithm can achieve relevant gains in terms of UAV endurance, without compromising the QoS offered by the flying network. As future work, we will consider the improvement of the EREP algorithm to other possible trajectories. Moreover, the improvement of EREP for positioning multiple UAVs and the use of an SNR margin with respect to the values obtained by means of the theoretical propagation models are research challenges worthy of being considered.

\section*{Acknowledgments}\label{sec:Acknowledgments}
This work is co-financed by the ERDF -- European Regional Development Fund through the Operational Programme for Competitiveness and Internationalisation -- COMPETE 2020 and the Lisboa2020 under the PORTUGAL 2020 Partnership Agreement, and through the Portuguese National Innovation Agency (ANI) as a part of the projects "CHIC: POCI-01-0247-FEDER-024498" and "5G: POCI-01-0247-FEDER-024539". The second author also thanks the funding from FCT under the PhD grant SFRH/BD/137255/2018.

\bibliography{references}

\begin{thebibliography}{10}
\providecommand \doibase [0]{http://dx.doi.org/}%

\bibitem{motlagh2017uav}
Motlagh NH, Bagaa M, Taleb T. {UAV}-{Based} {IoT} {Platform}: {A} {Crowd}
  {Surveillance} {Use} {Case}. {\it IEEE Communications Magazine} 2017\string;
  55(2)\string: 128--134.
\newblock \url{https://dx.doi.org/10.1109/MCOM.2017.1600587CM}.

\bibitem{coelho2018redefine}
Coelho A, Almeida EN, Silva P, Ruela J, Campos R, Ricardo M. RedeFINE:
  Centralized Routing for High-capacity Multi-hop Flying Networks. In: 2018
  IEEE 14th International Conference on Wireless and Mobile Computing,
  Networking and Communications (WiMob). IEEE; 2018\string: 75--82.
\newblock \url{https://dx.doi.org/10.1109/WiMOB.2018.8589098}.

\bibitem{almeida2018traffic}
Almeida EN, Campos R, Ricardo M. Traffic-aware multi-tier flying network:
  Network planning for throughput improvement. In: IEEE Wireless Communications
  and Networking Conference (WCNC). IEEE; 2018\string: 1--6.
\newblock \url{https://dx.doi.org/10.1109/WCNC.2018.8377408}.

\bibitem{zeng2016wireless}
Zeng Y, Zhang R, Lim TJ. Wireless communications with unmanned aerial vehicles:
  opportunities and challenges. {\it IEEE Communications Magazine} 2016\string;
  54(5)\string: 36-42.
\newblock \url{https://dx.doi.org/10.1109/MCOM.2016.7470933}.

\bibitem{Alzahrani2020}
Alzahrani B, Oubbati OS, Barnawi A, Atiquzzaman M, Alghazzawi D. UAV assistance
  paradigm: State-of-the-art in applications and challenges. {\it Journal of
  Network and Computer Applications} 2020\string: 102706.
\newblock \url{https://dx.doi.org/10.1016/j.jnca.2020.102706}.

\bibitem{Mozaffari2015}
Mozaffari M, Saad W, Bennis M, Debbah M. Drone small cells in the clouds:
  Design, deployment and performance analysis. In: 2015 IEEE Global
  Communications Conference (GLOBECOM). IEEE; 2015\string: 1--6.
\newblock \url{https://dx.doi.org/10.1109/GLOCOM.2015.7417609}.

\bibitem{Kalantari2017}
Kalantari E, Shakir MZ, Yanikomeroglu H, Yongacoglu A. Backhaul-aware robust 3D
  drone placement in 5G+ wireless networks. In: 2017 IEEE international
  conference on communications workshops (ICC workshops). IEEE; 2017\string:
  109--114.
\newblock \url{https://dx.doi.org/10.1109/ICCW.2017.7962642}.

\bibitem{Wu2018}
Wu Q, Zeng Y, Zhang R. Joint trajectory and communication design for multi-UAV
  enabled wireless networks. {\it IEEE Transactions on Wireless Communications}
  2018\string; 17(3)\string: 2109--2121.
\newblock \url{https://dx.doi.org/10.1109/TWC.2017.2789293}.

\bibitem{Arribas2019}
Arribas E, Mancuso V, Cholvi V. Coverage Optimization with a Dynamic Network of
  Drone Relays. {\it IEEE Transactions on Mobile Computing} 2020\string;
  19(10)\string: 2278-2298.
\newblock \url{https://dx.doi.org/10.1109/TMC.2019.2927335}.

\bibitem{Zeng2016}
Zeng Y, Zhang R, Lim TJ. Throughput maximization for UAV-enabled mobile
  relaying systems. {\it IEEE Transactions on Communications} 2016\string;
  64(12)\string: 4983--4996.
\newblock \url{https://dx.doi.org/10.1109/TCOMM.2016.2611512}.

\bibitem{Alzenad2017}
Alzenad M, El-Keyi A, Yanikomeroglu H. 3-D placement of an unmanned aerial
  vehicle base station for maximum coverage of users with different QoS
  requirements. {\it IEEE Wireless Communications Letters} 2017\string;
  7(1)\string: 38--41.
\newblock \url{https://dx.doi.org/10.1109/LWC.2017.2752161}.

\bibitem{Almeida2019}
{Almeida} EN, {Fernandes} K, {Andrade} F, {Silva} P, {Campos} R, {Ricardo} M. A
  Machine Learning Based Quality of Service Estimator for Aerial Wireless
  Networks. In: 2019 IEEE International Conference on Wireless and Mobile
  Computing, Networking and Communications (WiMob). IEEE; 2019\string: 1-6.
\newblock \url{https://dx.doi.org/10.1109/WiMOB.2019.8923217}.

\bibitem{Liu2019}
Liu L, Zhang S, Zhang R. CoMP in the sky: UAV placement and movement
  optimization for multi-user communications. {\it IEEE Transactions on
  Communications} 2019\string; 67(8)\string: 5645--5658.
\newblock \url{https://dx.doi.org/10.1109/TCOMM.2019.2907944}.

\bibitem{Almeida2020}
Almeida EN, Coelho A, Ruela J, Campos R, Ricardo M. Joint traffic-aware UAV
  placement and predictive routing for aerial networks. {\it Ad Hoc Networks}
  2021\string; 118\string: 102525.
\newblock \url{https://dx.doi.org/10.1016/j.adhoc.2021.102525}.

\bibitem{larsen2017optimal}
Larsen E, Landmark L, Kure {\O}. Optimal UAV relay positions in multi-rate
  networks. In: 2017 Wireless Days. IEEE; 2017\string: 8--14.
\newblock \url{https://dx.doi.org/10.1109/WD.2017.7918107}.

\bibitem{Zhong2019}
{Zhong} X, others . Deployment Optimization of UAV Relay for Malfunctioning
  Base Station: Model-Free Approaches. {\it IEEE Transactions on Vehicular
  Technology} 2019\string; 68(12)\string: 11971-11984.
\newblock \url{https://dx.doi.org/10.1109/TVT.2019.2947078}.

\bibitem{Zhong2020}
Zhong X, Guo Y, Li N, Chen Y. Joint Optimization of Relay Deployment, Channel
  Allocation, and Relay Assignment for UAVs-Aided D2D Networks. {\it IEEE/ACM
  Transactions on Networking} 2020\string; 28(2)\string: 804--817.
\newblock \url{https://dx.doi.org/10.1109/TNET.2020.2970744}.

\bibitem{Li2015}
Li K, Ni W, Wang X, Liu RP, Kanhere SS, Jha S. EPLA: Energy-balancing packets
  scheduling for airborne relaying networks. In: 2015 IEEE International
  Conference on Communications (ICC). IEEE; 2015\string: 6246--6251.
\newblock \url{https://dx.doi.org/10.1109/ICC.2015.7249319}.

\bibitem{Li2016}
Li K, Ni W, Wang X, Liu RP, Kanhere SS, Jha S. Energy-efficient cooperative
  relaying for unmanned aerial vehicles. {\it IEEE Transactions on Mobile
  Computing} 2015\string; 15(6)\string: 1377--1386.
\newblock \url{https://dx.doi.org/10.1109/TMC.2015.2467381}.

\bibitem{Li2019}
Li K, Voicu RC, Kanhere SS, Ni W, Tovar E. Energy efficient legitimate wireless
  surveillance of UAV communications. {\it IEEE Transactions on Vehicular
  Technology} 2019\string; 68(3)\string: 2283--2293.
\newblock \url{https://dx.doi.org/10.1109/TVT.2019.2890999}.

\bibitem{Qin2019trajectory}
Qin Z, Dong C, Wang H, et al. Trajectory Planning for Data Collection of
  Energy-Constrained Heterogeneous UAVs. {\it Sensors} 2019\string;
  19(22)\string: 4884.
\newblock \url{https://dx.doi.org/10.3390/s19224884}.

\bibitem{zeng2019energy}
Zeng Y, Xu J, Zhang R. Energy minimization for wireless communication with
  rotary-wing UAV. {\it IEEE Transactions on Wireless Communications}
  2019\string; 18(4)\string: 2329--2345.
\newblock \url{https://dx.doi.org/10.1109/TWC.2019.2902559}.

\bibitem{coelho2019traffic}
Coelho A, Fontes H, Campos R, Ricardo M. Traffic-aware gateway placement for
  high-capacity flying networks. In: 2021 IEEE 93rd Vehicular Technology
  Conference (VTC2021-Spring). IEEE; 2021\string: 1--6.
\newblock \url{https://dx.doi.org/10.1109/VTC2021-Spring51267.2021.9448966}.

\bibitem{Oubbati2019}
Oubbati OS, Mozaffari M, Chaib N, Lorenz P, Atiquzzaman M, Jamalipour A. ECaD:
  Energy-efficient routing in flying ad hoc networks. {\it International
  Journal of Communication Systems} 2019\string; 32(18)\string: e4156.
\newblock \url{https://dx.doi.org/10.1002/dac.4156}.

\bibitem{Aadil2018}
Aadil F, Raza A, Khan MF, Maqsood M, Mehmood I, Rho S. Energy aware
  cluster-based routing in flying ad-hoc networks. {\it Sensors} 2018\string;
  18(5)\string: 1413.
\newblock \url{https://dx.doi.org/10.3390/s18051413}.

\bibitem{Alshabtat2010}
Alshabtat AI, Dong L, Li J, Yang F. Low latency routing algorithm for unmanned
  aerial vehicles ad-hoc networks. {\it International Journal of Electrical and
  Computer Engineering} 2010\string; 6(1)\string: 48--54.
\newblock \url{https://dx.doi.org/10.5281/zenodo.1061573}.

\bibitem{Fan2012}
Fan Q, Fan J, Li J, Wang X. A multi-hop energy-efficient sleeping MAC protocol
  based on TDMA scheduling for wireless mesh sensor networks. {\it Journal of
  Networks} 2012\string; 7(9)\string: 1355.
\newblock
  \url{https://citeseerx.ist.psu.edu/viewdoc/summary?doi=10.1.1.369.7535}.

\bibitem{Yang2004}
Yang X, Vaidya NH. A wakeup scheme for sensor networks: Achieving balance
  between energy saving and end-to-end delay. In: 10th IEEE Real-Time and
  Embedded Technology and Applications Symposium, 2004. IEEE; 2004\string:
  19--26.
\newblock \url{https://dx.doi.org/10.1109/RTTAS.2004.1317245}.

\bibitem{Mozaffari2017}
Mozaffari M, Saad W, Bennis M, Debbah M. Wireless communication using unmanned
  aerial vehicles (UAVs): Optimal transport theory for hover time optimization.
  {\it IEEE Transactions on Wireless Communications} 2017\string;
  16(12)\string: 8052--8066.
\newblock \url{https://dx.doi.org/10.1109/TWC.2017.2756644}.

\bibitem{Halperin2010}
Halperin D, Greenstein B, Sheth A, Wetherall D. Demystifying 802.11n Power
  Consumption. In: HotPower'10. Proceedings of the 2010 International
  Conference on Power Aware Computing and Systems. USENIX Association; 2010;
  USA\string: 1-5.
\newblock
  \url{https://citeseerx.ist.psu.edu/viewdoc/summary?doi=10.1.1.173.7044}.

\bibitem{Sousa2018}
{Sousa} F, {Dias} J, {Ribeiro} F, {Campos} R, {Ricardo} M. Green Wireless Video
  Sensor Networks Using Low Power Out-of-Band Signalling. {\it IEEE Access}
  2018\string; 6\string: 30024-30038.
\newblock \url{https://dx.doi.org/10.1109/ACCESS.2018.2841821}.

\bibitem{Zeng2017}
Zeng Y, Zhang R. Energy-efficient UAV communication with trajectory
  optimization. {\it IEEE Transactions on Wireless Communications} 2017\string;
  16(6)\string: 3747--3760.
\newblock \url{https://dx.doi.org/10.1109/TWC.2017.2688328}.

\bibitem{Xiaowei2019}
Li X, Yao H, Wang J, Xu X, Jiang C, Hanzo L. A Near-Optimal UAV-Aided Radio
  Coverage Strategy for Dense Urban Areas. {\it IEEE Transactions on Vehicular
  Technology} 2019\string; 68(9)\string: 9098-9109.
\newblock \url{https://dx.doi.org/10.1109/TVT.2019.2927425}.

\bibitem{Xiaowei2019IoT}
Li X, Yao H, Wang J, Wu S, Jiang C, Qian Y. Rechargeable Multi-UAV Aided
  Seamless Coverage for QoS-Guaranteed IoT Networks. {\it IEEE Internet of
  Things Journal} 2019\string; 6(6)\string: 10902-10914.
\newblock \url{https://dx.doi.org/10.1109/JIOT.2019.2943147}.

\bibitem{demir2019energy}
Demir U, {\.I}pek M{\c{C}}, Toker C, Ekici {\"O}. Energy-Efficient Rotary-Wing
  UAV Deployment Under Flight Dynamics and QoS Constraints. In: 2019 IEEE
  International Black Sea Conference on Communications and Networking
  (BlackSeaCom). IEEE; 2019\string: 1--5.
\newblock \url{https://dx.doi.org/10.1109/BlackSeaCom.2019.8812816}.

\bibitem{Jingjing2019}
Wang J, Jiang C, Wei Z, Pan C, Zhang H, Ren Y. Joint UAV Hovering Altitude and
  Power Control for Space-Air-Ground IoT Networks. {\it IEEE Internet of Things
  Journal} 2019\string; 6(2)\string: 1741-1753.
\newblock \url{https://dx.doi.org/10.1109/JIOT.2018.2875493}.

\bibitem{Babu2020_1}
Babu N, Papadias CB, Popovski P. Energy-Efficient 3D Deployment of Aerial
  Access Points in a UAV Communication System. {\it IEEE Communications
  Letters} 2020.
\newblock \url{https://dx.doi.org/10.1109/LCOMM.2020.3017559}.

\bibitem{Babu2020_2}
{Babu} N, {Ntougias} K, {Papadias} CB, {Popovski} P. Energy Efficient Altitude
  Optimization of an Aerial Access Point. In: 2020 IEEE 31st Annual
  International Symposium on Personal, Indoor and Mobile Radio Communications.
  IEEE; 2020\string: 1-7.
\newblock \url{https://dx.doi.org/10.1109/PIMRC48278.2020.9217265}.

\bibitem{friis1946note}
Friis HT. A note on a simple transmission formula. {\it Proceedings of the
  I.R.E. and Waves and Electrons} 1946\string; 34(5)\string: 254--256.
\newblock
  \url{https://capmimo.ece.wisc.edu/capmimo_papers/friis_original_1946.pdf}.

\bibitem{Khuwaja2018}
Khuwaja AA, Chen Y, Zhao N, Alouini MS, Dobbins P. A survey of channel modeling
  for UAV communications. {\it IEEE Communications Surveys \& Tutorials}
  2018\string; 20(4)\string: 2804--2821.
\newblock \url{https://dx.doi.org/10.1109/COMST.2018.2856587}.

\bibitem{Zhan2018}
Zhan C, Zeng Y, Zhang R. Trajectory design for distributed estimation in
  UAV-enabled wireless sensor network. {\it IEEE Transactions on Vehicular
  Technology} 2018\string; 67(10)\string: 10155--10159.
\newblock \url{https://dx.doi.org/10.1109/TVT.2018.2859450}.

\bibitem{Samir2019}
Samir M, Sharafeddine S, Assi CM, Nguyen TM, Ghrayeb A. UAV trajectory planning
  for data collection from time-constrained IoT devices. {\it IEEE Transactions
  on Wireless Communications} 2019\string; 19(1)\string: 34--46.
\newblock \url{https://dx.doi.org/10.1109/TWC.2019.2940447}.

\bibitem{Zorbas2016}
Zorbas D, Pugliese LDP, Razafindralambo T, Guerriero F. Optimal drone placement
  and cost-efficient target coverage. {\it Journal of Network and Computer
  Applications} 2016\string; 75\string: 16--31.
\newblock \url{https://dx.doi.org/10.1016/j.jnca.2016.08.009}.

\bibitem{Abeywickrama2018}
Abeywickrama HV, Jayawickrama BA, He Y, Dutkiewicz E. Comprehensive energy
  consumption model for unmanned aerial vehicles, based on empirical studies of
  battery performance. {\it IEEE Access} 2018\string; 6\string: 58383--58394.
\newblock \url{https://dx.doi.org/10.1109/ACCESS.2018.2875040}.

\bibitem{MCSIndexTable:online}
Pros W. MCS Index chart - 802.11ac - VHT.
  \url{https://www.wlanpros.com/mcs-index-charts/};  2021.
\newblock (Accessed on 02/12/2021).

\bibitem{rodrigues2020}
Rodrigues H. Gateway Positioning in Flying Networks. Master's thesis. Faculdade
  de Engenharia, Universidade do Porto, Portugal.
  \url{https://hdl.handle.net/10216/132922}:   2020.

\bibitem{UAVPowerSimulator}
UAV Power Simulator. \url{https://github.com/rhugo97/UAV-Power-Simulator};
  2020.
\newblock (Accessed on 02/12/2021).

\bibitem{aschenbruck2010bonnmotion}
Aschenbruck N, Ernst R, Gerhards-Padilla E, Schwamborn M. BonnMotion: a
  mobility scenario generation and analysis tool. In: Proceedings of the 3rd
  international ICST conference on simulation tools and techniques. ICST
  (Institute for Computer Sciences, Social-Informatics and Telecommunications
  Engineering); 2010\string: 1--10.
\newblock \url{https://dx.doi.org/10.4108/ICST.SIMUTOOLS2010.8684}.

\bibitem{ns3Simulator}
NS-3 . {Network Simulator}. \url{https://www.nsnam.org/};  2021.
\newblock (Accessed on 02/12/2021).

\bibitem{queue}
Nichols K, Jacobson V. {\it Controlling queue delay}.
\newblock Queue, vol. 10 .
\newblock 2012.

\bibitem{ns3Codel}
CoDel queue disc. \url{https://www.nsnam.org/docs/models/html/codel.html};
  2021.
\newblock (Accessed on 02/12/2021).

\bibitem{gettys2011}
Gettys J, Nichols K. Bufferbloat: Dark Buffers in the Internet: Networks
  without Effective AQM May Again Be Vulnerable to Congestion Collapse.. {\it
  Queue} 2011\string; 9(11)\string: 40–54.
\newblock \url{https://dx.doi.org/10.1145/2063166.2071893}.

\end{thebibliography}

\end{document}